\documentclass[twocolumn,english,groupedaddress,superscriptaddress,aps,prl,10pt,floatfix,nobibnotes]{revtex4-2}

\usepackage[T1]{fontenc}
\usepackage[utf8]{inputenc}
\usepackage{amsmath,amsthm,mathtools,amssymb}
\usepackage{graphicx}
\usepackage{dsfont}
\usepackage{url}

\usepackage[breaklinks]{hyperref}
\usepackage{orcidlink}
\usepackage{eurosym}
\usepackage{siunitx}
\DeclareSIUnit{\EUR}{\text{\euro}}

\renewcommand{\figurename}{\textbf{Figure}}

\newcounter{notecounter}
\newcounter{methodcounter}

\begin{document}

\title{Understanding the European energy crisis through structural causal models}

\author{Sarah Schreyer \orcidlink{0009-0003-5866-9233}}\thanks{These authors contributed equally.}
\affiliation{Institute of Climate and Energy Systems: Energy Systems Engineering (ICE-1), Forschungszentrum J\"ulich, 52428 J\"ulich, Germany}
\affiliation{Institute for Theoretical Physics, University of Cologne, 50937 K\"oln, Germany}

\author{Anton Tausendfreund \orcidlink{0009-0001-9148-3188}}\thanks{These authors contributed equally.}
\affiliation{Institute of Climate and Energy Systems: Energy Systems Engineering (ICE-1), Forschungszentrum J\"ulich, 52428 J\"ulich, Germany}
\affiliation{Institute for Theoretical Physics, University of Cologne, 50937 K\"oln, Germany}

\author{Florian Immig \orcidlink{0009-0006-6300-8773}}
\thanks{These authors contributed equally.}
\affiliation{Institute for Automation and Applied Informatics, Karlsruhe Institute of Technology, 76344 Eggenstein-Leopoldshafen, Germany}

\author{Ulrich Oberhofer \orcidlink{0009-0001-2766-5796}}
\thanks{These authors contributed equally.}
\affiliation{Institute for Automation and Applied Informatics, Karlsruhe Institute of Technology, 76344 Eggenstein-Leopoldshafen, Germany}

\author{Julius Trebbien \orcidlink{0009-0003-5866-9233}}
\affiliation{Institute of Climate and Energy Systems: Energy Systems Engineering (ICE-1), Forschungszentrum J\"ulich, 52428 J\"ulich, Germany}
\affiliation{Institute for Theoretical Physics, University of Cologne, 50937 K\"oln, Germany}

\author{Aaron Praktiknjo
\orcidlink{0000-0002-2151-4241}
}
\affiliation{Chair for Energy System Economics, Institute for Future Energy Consumer Needs and Behavior (FCN), E.ON Energy Research Center, RWTH Aachen University, 52074 Aachen, Germany}
\affiliation{JARA-ENERGY, 52074 Aachen, Germany}

\author{Benjamin Schäfer \orcidlink{0000-0003-1607-9748}}
\thanks{Corresponding author: Mail address: 
\href{benjamin.schaefer@kit.edu}{benjamin.schaefer@kit.edu}}
\affiliation{Institute for Automation and Applied Informatics, Karlsruhe Institute of Technology, 76344 Eggenstein-Leopoldshafen, Germany}

\author{Dirk Witthaut \orcidlink{0000-0002-3623-5341}}
\thanks{Corresponding author: Mail address: \href{d.witthaut@fz-juelich.de}{d.witthaut@fz-juelich.de}}
\affiliation{Institute of Climate and Energy Systems: Energy Systems Engineering (ICE-1), Forschungszentrum J\"ulich, 52428 J\"ulich, Germany}
\affiliation{Institute for Theoretical Physics, University of Cologne, 50937 K\"oln, Germany}
\affiliation{JARA-ENERGY, 52074 Aachen, Germany}

\begin{abstract}
Natural gas supplies in Europe were disrupted and energy prices soared in the context of Russia’s invasion of Ukraine. Electricity prices in France experienced the largest relative increase among European countries, even though the share of natural gas in the electricity mix is small compared to its neighbours. In this article, we demonstrate the importance of causal statistical methods and propose causal graphs to investigate the French and Spanish electricity markets and pinpoint key influencing factors on electricity prices and net exports. We demonstrate that a causal approach resolves paradoxical results of simple correlation studies and enables a quantitative analysis of indirect causal effects and what-if scenarios. We introduce a linear structural causal model as well as non-linear tree-based machine learning combined with Shapley Flow values. The models elucidate the interplay of gas prices and the unavailability of nuclear power plants during the energy crisis as the high unavailability made France dependent on imports.
\end{abstract}

\maketitle

\section*{Introduction}

\begin{figure*}[ht]
    \centering
    \includegraphics[width=\linewidth]{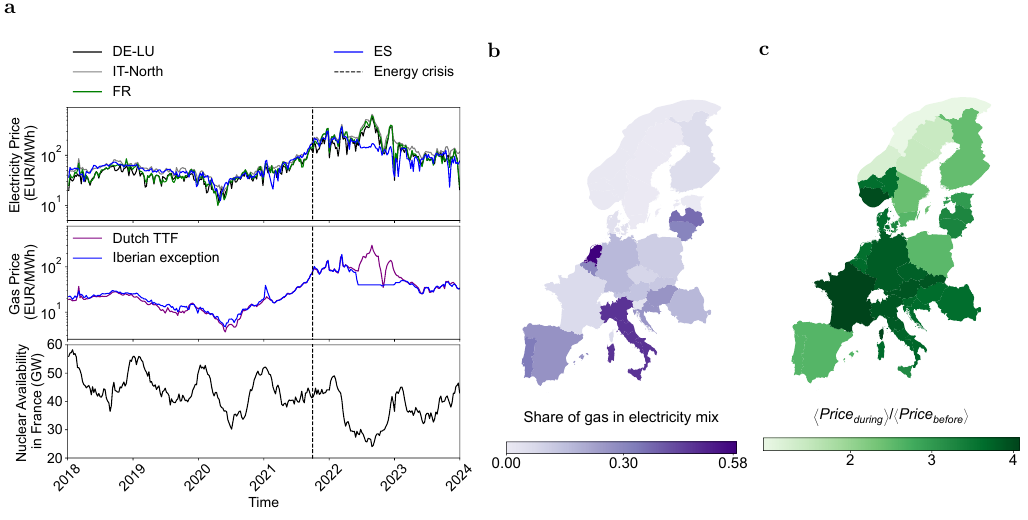}           
    \caption{
    \textbf{The European energy crisis.}
    \textbf{a,} 
    Energy prices soared in connection to the Russian invasion of Ukraine.
    Here, we show gas prices from the Dutch TTF daily future market as well as capped gas prices under the Iberian exception and electricity prices from the day-ahead spot markets in the bidding zones France (FR), Germany-Luxembourg (DE-LU), Spain (ES) and Italy North (IT-North) averaged weekly on a logarithmic scale. In France, the available nuclear power dropped to very low values during the crisis due to urgent revisions. Prices increased well before the invasion; here we define October 1 (dashed line) as the beginning of the energy crisis.   
    \textbf{b,} The share of natural gas in the electricity mix differs strongly between European countries. Data were taken from before the crisis (year 2020).
    \textbf{c,} Relative increase of electricity market prices in the different bidding zones. Prices in France and Southern Norway increased by a factor of approximately four, although the contribution of natural gas is negligible in these countries. 
    Raw data has been obtained from LSEG~\cite{thomsonreuters}, ENTSO-E \cite{entsoeElectricityMarket} and Eurostat \cite{europa} (see Methods for details). ENTSO-E bidding zone geometries were accessed using the entsoe-py package \cite{githubGitHubEnergieIDentsoepy}. 
    }
    \label{fig:energy crisis}
\end{figure*}

The European energy system is dependent on imports of natural gas and other fossil fuels~\cite{chalvatzis2017energy,pedersen2022long}. Natural gas supplies were disrupted in the context of the Russian invasion of Ukraine, as major pipelines pass through Ukraine, and Russia shut down the NordStream1 pipeline in July 2022~\cite{goodell2023global}. Market prices for natural gas and electricity spiked across Europe, with gas prices peaking at $330 \, \si{EUR}/\si{MWh}$ (Fig.~\ref{fig:energy crisis}a). 
As a result, energy prices and energy security have become dominant issues in the political debate across Europe~\cite{goldthau2022energy}. Many countries have lowered energy taxes~\cite{gars2022effect} and increased efforts to reduce consumption~\cite{ruhnau2023natural}. The European Union adopted measures to reduce import dependency from Russia and promote the security of supply~\cite{eu2022repowereu, gas_source}.

Notably, electricity markets were affected quite differently during the crisis~\cite{trebbien2024patterns}: 
The relative price increase was lowest in Poland and the northern parts of Norway, while a strong relative increase in prices was observed in Norway's southernmost bidding zone NO2 and in France.
In both regions electricity market prices increased by almost a factor of four (Fig.~\ref{fig:energy crisis}b). This is very surprising as natural gas plays only a limited role in electricity generation in both countries. In France, baseload generation relies primarily on nuclear power, while in Norway, it is mainly covered by hydroelectric power~\cite{entsoeElectricityMarket}. So what caused the drastic increase in electricity prices? Why were France and Norway affected so strongly, despite comparatively low natural gas use in their electricity systems?

In general, debates about the future energy system remain highly controversial, in particular with regard to the role of nuclear power plants~\cite{jenkins2018getting,haywood2023investing}
and renewable energy sources~\cite{pahle2022safeguarding,brown2021decreasing}.
At their core, these debates concern the consequences of concrete interventions, such as capacity expansion, plant closures, or market regulation. Assessing such consequences requires empirical evidence on how structural changes propagate through electricity systems and affect market outcomes. The increasing availability of detailed operational and market data~\cite{entsoeElectricityMarket} enables systematic analysis of such complex interactions~\cite{khodayar2020deep,strielkowski2023prospects}. However, while predictive models and correlation analyses can describe patterns, they do not by themselves identify the effects of interventions. To inform system design and policy choices, causal explanations are required~\cite{scholkopf2021toward}.

In this article, we show the importance of a causal analysis of electricity market data in the context of the European energy crisis. We present two approaches based on causal graphs: linear structural causal models~\cite{pearl2016causal} and non-linear tree-based machine learning combined with Shapley Flow values~\cite{wang2021shapley}. We focus primarily on the French day-ahead (DA) electricity spot market and disentangle the role of rising natural gas prices and problems with the nuclear fleet~\cite{banque2023energy}, which occurred coincidentally. 
Given France’s interconnection with the Iberian Peninsula, we additionally examine cross-border effects on prices in the Spanish day-ahead (DA) electricity spot market.
Furthermore, we analyse indirect causal effects and quantify how weather effects, such as river flow and temperature, affected load and nuclear availability and thereby prices and export capacities in France. Building upon this analysis, we use our model to examine selected what-if cases.
Although the onset of the energy crisis spans a period of time, we fix October 1, 2021, as a reference date and compare the periods before and after (see Methods).

\section*{Results}

\subsection*{Correlation and causality in electricity markets}

The French electricity system is mainly based on nuclear power, with a share of more than 65\% in the yearly electricity mix before the energy crisis~\cite{entsoeElectricityMarket}.
In 2022 and 2023, several plants were out of operation due to delayed maintenance, corrosion problems or lack of cooling water~\cite{banque2023energy}. As a result, nuclear availability reached historically low levels during the energy crisis. Approximately half of the nuclear power plants were not available in summer 2022, and the availability occasionally dropped below $30 \, \si{GW}$ (Fig.~\ref{fig:energy crisis}a). It is therefore plausible that problems with the nuclear fleet were a major contributor to rising electricity prices. These problems were specific to the French electricity system and may explain why France was more affected than neighbouring countries.

\begin{figure*}[ht]
    \centering
    \includegraphics[width=\linewidth]{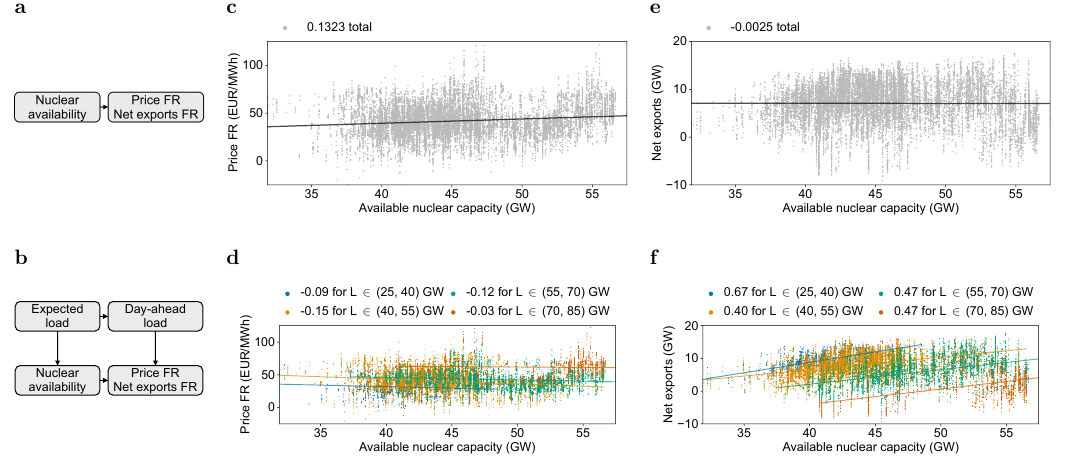}       
    \caption{
    \textbf{Simpson's paradox in the French electricity market.}
    \textbf{a,} Graphical representation of the relation of the available nuclear capacity and the electricity market price, where causal effects are represented by arrows. Here, we ignore confounding effects. 
    \textbf{b,} In a modified model, we include confounding via the expected load, that affects both the actual day-ahead load and the nuclear availability via outage planning.
    \textbf{c,} Relation of the available nuclear capacity and the electricity market price. The price seemingly increases with the capacity, counterintuitive to economic laws. The Pearson correlation is positive (see legend), a linear fit (solid line) is shown to facilitate interpretation.
    \textbf{d,} The paradox is resolved if we adjust for confounding by splitting the time series into four groups according to the load $L$. The correlation coefficient is negative for all groups (see legend).
    \textbf{e,} A similar pattern is observed for the net exports, i.e.~the difference of electricity exports and imports. For the entire dataset, the correlation is negative (see legend) contrary to expectations. 
    \textbf{f,} Splitting the data to four group according to the load $L$ yields a positive correlation as expected (see legend).
    Here, we use data from 1 April 2018 to 1 April 2020 only to exclude impacts of the energy crisis. Every data point corresponds to one hour. 
    }
    \label{fig:simpsons paradox}
\end{figure*}

However, empirical validation of this hypothesis is not straightforward. A naive data analysis shows that the correlation between electricity market prices and nuclear availability is positive (Fig.~\ref{fig:simpsons paradox}b). That is, the price appears to increase with the availability of nuclear power and thus with higher supply, counterintuitive to economic laws. To resolve this paradox, we need to distinguish causal effects from mere correlations. The French electricity system is subject to strong seasonal effects that lead to confounding: The load, and thus the demand, is significantly higher in the winter than in the summer and high electricity prices typically occur in winter
(see Supplementary Fig.~\ref{fig:na_seasonality} and Supplementary Note~1). Hence, revisions take place mostly in the summer where both demand and electricity prices are lowest and thus the opportunity costs for unavailability are lowest.
Therefore, the observed positive correlation between electricity price and nuclear availability does not necessarily indicate a causal effect.

This situation is illustrated in Fig.~\ref{fig:simpsons paradox}a,b, where arrows represent hypothesized causal effects. The relationship between nuclear availability and electricity market prices is governed by two paths: (i) a direct causal effect of availability on prices, consistent with the law of supply, whereby higher available capacity tends to reduce prices, and (ii) an indirect path via expected and realized load. Expected load influences planned outages, as maintenance is typically scheduled during periods of forecasted low demand and low prices. At the same time, expected load strongly predicts the realized day-ahead load, which in turn affects prices. Expected load therefore acts as a confounder: periods of high expected load are associated with both higher nuclear availability and higher prices. This induces a positive correlation between availability and price, even though the direct causal effect of additional nuclear supply would reduce prices.

To estimate the causal effect of nuclear availability on the electricity market prices, the spurious path through load needs to be accounted for. One transparent way to do so is to divide the data into groups with similar realized load $L$. Within each group, we observe a negative relationship between electricity market price and nuclear availability (Fig.~\ref{fig:simpsons paradox}b). Once the confounding effect of load is controlled for, the relationship aligns with the law of supply and demand.

A similar pattern is observed for French net exports, defined as the difference between exports and imports. The raw data shows a negative correlation to the available nuclear capacity. After accounting for the confounding factor, we observe  a positive correlation as expected (Fig.~\ref{fig:simpsons paradox}c).

The above example shows that advanced statistical methods are needed in the empirical analysis of electricity markets. In particular, we need methods of causal inference to deal with confounders and to distinguish correlations from causal effects. In what follows, we turn to structural causal models for quantitative analysis. 

\subsection*{A structural causal model for electricity market prices and exports}

\begin{figure*}[ht]
    \centering
    \includegraphics[width=\textwidth]{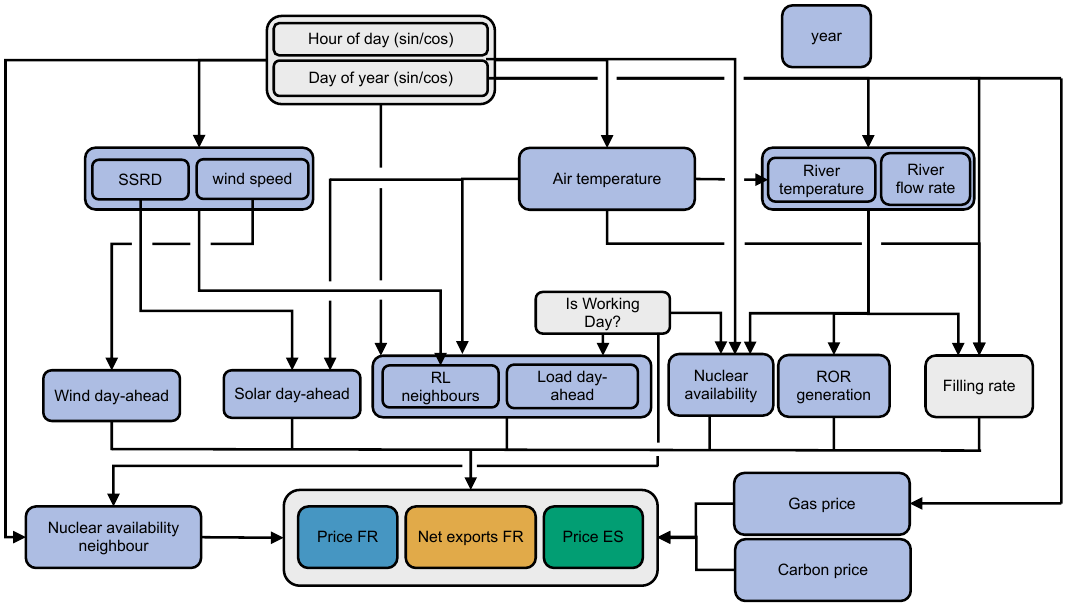}
    \caption{\textbf{Structural causal model (SCM) for the French and Spanish electricity market.}
    We propose a causal graph for the analysis of electricity prices and net exports. Boxes represent calendrical, meteorological and energy system variables included in our analysis (see main text for details). Variables with the same neighbours in the causal graph are grouped for the sake of clarity. Arrows indicate potential causal interactions between two variables. Most variables depend on the year; these are marked by a blue background for clarity.
    We use the abbreviations ROR for run-of-river hydropower and RL for the residual load, i.e.~load minus non-dispatchable generation.
    }
    \label{fig:structural causal model}
\end{figure*}

We propose a structural causal model (SCM) to analyse influences on the electricity market prices in France and Spain and net exports of France during the energy crisis (Fig.~\ref{fig:structural causal model}). The starting point is the formulation of a causal graph, where each variable of interest $X_i$ is described by a node~\cite{pearl2016causal}. A potential causal interaction of two variables is described by a directed edge. We construct this graph for prices and net exports based on domain knowledge in the following.

The electricity market price and net exports are mainly affected by load, renewable generation, availability of other plants or filling rate of hydro storage, as well as prices for natural gas and carbon emissions~\cite{praktiknjo2016renewable,trebbien2023understanding}. To capture cross-border effects, we also include the residual load and nuclear availability of neighbouring countries; this will be discussed in detail below. The feature variables themselves depend on calendrical and meteorological variables. 

In an SCM, each variable $X_i$ is described by a structural equation $X_i := f_i(PA_i, U_i)$, where $PA_i$ is the set of parents of $X_i$ in the causal graph and $U_i$ describes unobserved variables. Restricting ourselves to linear equations enables a straightforward interpretation of the results. The structural equations simplify to
\begin{align}
    X_i := \sum_{j \in {\rm parents}(i)} c_{ij} X_j + U_i ,
    \label{eq:structural equation}
\end{align}
where the structural coefficient $c_{ij}$ describes the direct causal effect of the variable $X_j$ on the variable $X_i$. We emphasise that although we use regression to infer structural coefficients, these coefficients are not identical to ordinary regression coefficients. Structural coefficients assume causation, regression coefficients do not~\cite{pearl2016causal}.
The linear formulation enables a transparent and interpretable quantification of direct causal effects. To assess potential non-linearities and indirect effects, we complement this analysis with a Shapley Flow approach below.

\begin{figure*}[ht]
    \centering
    \includegraphics[width=\linewidth]{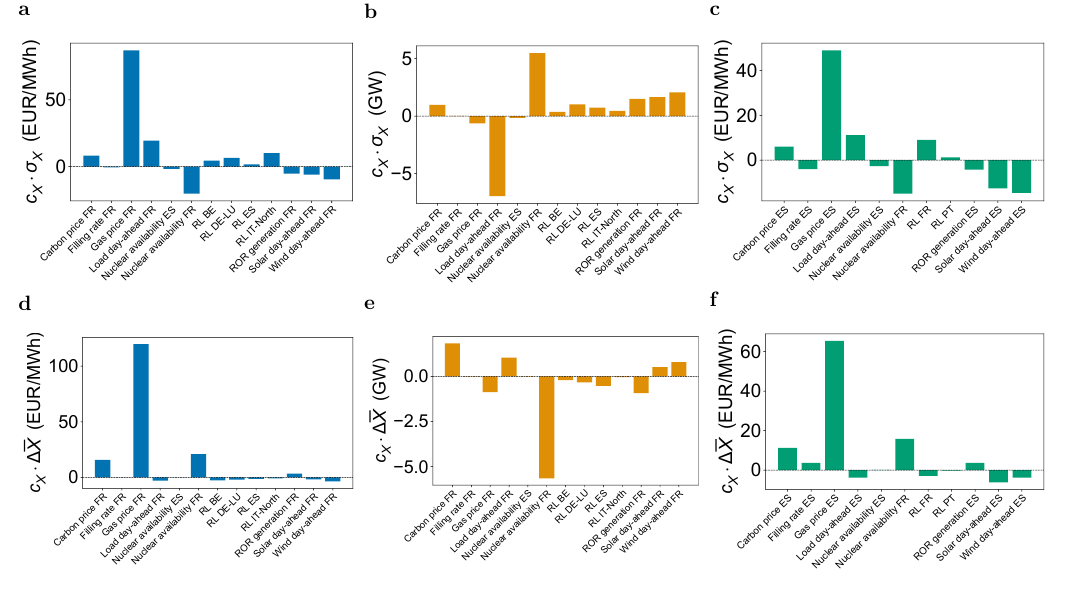}       
    \caption{
    \textbf{Results of the SCMs for the French and Spanish electricity market.}
    Results are shown for \textbf{a,d} the electricity market prices of France, \textbf{b,e} the net exports of France and \textbf{c,f} the electricity market prices of Spain.
    \textbf{a,b,c,} Structural coefficients $c_{ij}$ of the SCM normalized by the standard deviation $\sigma_j$ of the predictor variable to enable comparability. 
    \textbf{d,e,f,} Structural coefficients $c_{ij}$ multiplied by the average difference of the variable before and during the energy crisis $\Delta\overline{X_j}$ as defined in Eq.~\eqref{eq:cDeltaX}.
    This quantity measures the impact of changes in the predictor variable on the respective target. For the electricity market prices of France and Spain, the increase of the gas price has the strongest impact with more than $+110 \, \si{EUR}/\si{MWh}$ and $+65 \, \si{EUR}/\si{MWh}$, respectively. For the net exports of France, the decrease of the nuclear availability had the strongest impact with $-5.6 \, \si{GW}$.
    }
    \label{fig:structural_coefficients}
\end{figure*}

We fit the proposed linear SCM to French and Spanish data covering the period 2018--2023. For electricity market prices, the models achieve $R^2$ scores of $0.92$ and $0.91$, respectively (Supplementary Tab.~\ref{tab:model_performance}), indicating strong explanatory power. In contrast, the $R^2$ values for intermediate variables vary substantially (Supplementary Fig.~\ref{fig:R2 scores}). Lower values suggest either limitations of the linear approximation or the presence of relevant exogenous drivers not included in the model. For example, wind power generation is largely determined by synoptic meteorological conditions that are not explicitly modelled.

The resulting structural coefficients are shown in Fig.~\ref{fig:structural_coefficients}, scaled by the respective standard deviations; unscaled coefficients are provided in Supplementary Fig.~\ref{fig:coefficients}.
To capture the role of each variable during the energy crisis, we also plot the quantity
\begin{align}
    c_{ij} \Delta \bar X_{j} = 
    c_{ij} \left( \left\langle X_j \right\rangle_{\rm during} - \left\langle X_j \right\rangle_{\rm before} \right),
    \label{eq:cDeltaX}
\end{align}
where $\left\langle X_j \right\rangle_{\text{ before/during}}$ denotes the average of the respective variable before and during the crisis. This variable quantifies how a change in the variable $X_j$ during the energy crisis causally affected the variable $X_i$ on average.

We find that the natural gas price is the most important variable affecting electricity market prices in both France and Spain. In the following, we focus on France and return to the Spanish market when discussing cross-border effects.

In the model for French electricity prices, nuclear availability ranks second in importance, albeit at a much smaller scale than the gas price. While its direct effect on prices is limited, it plays an important structural role by shaping France’s position in cross-border trade, as discussed below.

The importance of the gas price is not generally unexpected: In many electricity markets, gas-fired power plants are the last to enter the market according to the merit order principle~\cite{sensfuss2008merit}. Their bids therefore determine the clearing price of the electricity spot market. However, the result is surprising in this context as natural gas plays a limited role in the French electricity market (Fig.~\ref{fig:energy crisis}b). In particular, it does not explain why France was affected stronger than its neighbours during the energy crisis.

France has been a net exporter of electricity for decades. With large parts of its nuclear fleet under revision, France could no longer maintain these exports and became a net importer in 2022~\cite{trebbien2024patterns}. In the same year, the share of natural gas in the electricity mix increased strongly (Supplementary Fig.~\ref{fig:gas generation}).  
The crucial role of nuclear availability for cross-border trading is confirmed by our structural causal model (Fig.~\ref{fig:structural_coefficients}d). We find that the difference in nuclear availability before and during the crisis essentially determines the sharp fall in France's net exports. 
In periods of high nuclear availability, domestic generation typically exceeds demand and France exports electricity. When nuclear availability is low, imports and gas-fired power plants are often needed to meet demand and therefore play a decisive role in setting the market clearing price.

Cross-border trade is closely linked to differences in electricity market prices. The European market coupling algorithm EUPHEMIA will generally increase exports to neighbouring countries when prices are lower and transmission capacity is available~\cite{all_nemo_committee_euphemia_2020,trebbien2024patterns}. The sharp fall in French net exports is thus associated with a reversal of price differences with respect to Germany, Belgium and Spain (Supplementary Fig.~\ref{fig:pricesync}). 
Reduced nuclear availability strengthened price coupling with neighbouring markets and increased the relevance of gas-based price formation for the French market.

The sharp rise in French electricity market prices can thus be understood as the interaction of two factors~\cite{banque2023energy}. Reduced nuclear availability increased France’s reliance on imports and gas-fired generation, while rising gas prices elevated the marginal cost of electricity across interconnected European markets.
%
It must be emphasised that the situation in the French electricity system could have been much more severe without the option to compensate lacking domestic generation by imports. European cooperation contributed to France's security of supply.

\subsection*{Cross-border effects}

We now turn to cross-border effects and examine how developments in one national electricity system influence electricity market prices in the other. In the SCMs for France and Spain, we therefore include the nuclear availability of the neighbouring country. Furthermore, we include the residual load, i.e.~the difference between load and renewable generation, for all neighbouring bidding zones that are part of the single day-ahead market coupling~\cite{entso-e_single}.

For the Spanish electricity market price, the most important driver during the energy crisis is again the natural gas price as discussed above. The second most important factor is the nuclear availability of France (Fig.~\ref{fig:structural_coefficients}c). This highlights how relevant cross-border effects can be for electricity price formation.

In contrast, the effect of Spanish nuclear availability on French electricity market prices and net exports is negligible in aggregate terms. Although the estimated cross-border structural coefficients are of similar magnitude (Supplementary Fig.~\ref{fig:coefficients}), the aggregate impact differs substantially. Spain’s nuclear generation capacity is considerably smaller and did not experience a comparable reduction during the crisis. Consequently, variations in Spanish nuclear availability were too limited to generate a sizable effect on the French market.

Cross-border influences are also reflected in residual load variables. In the model for Spanish electricity market prices, the residual load of France ranks sixth in terms of the scaled structural coefficient $c_X \sigma_X$ (Fig.~\ref{fig:structural_coefficients}\textbf{c}). Notably, its contribution exceeds that of Spanish run-of-river generation and the carbon price. In contrast, Spanish residual load plays only a minor role in the French model. This asymmetry mirrors the relative size of the two systems and France’s traditional role as a major exporter of electricity.

Overall, the results show that shocks to the French nuclear generation fleet propagated to Spain, demonstrating how interconnected electricity markets transmit price effects across borders.

\subsection*{Quantifying non-linear and indirect effects via Shapley Flow}

\begin{figure*}[ht] 
    \includegraphics[width=\linewidth]{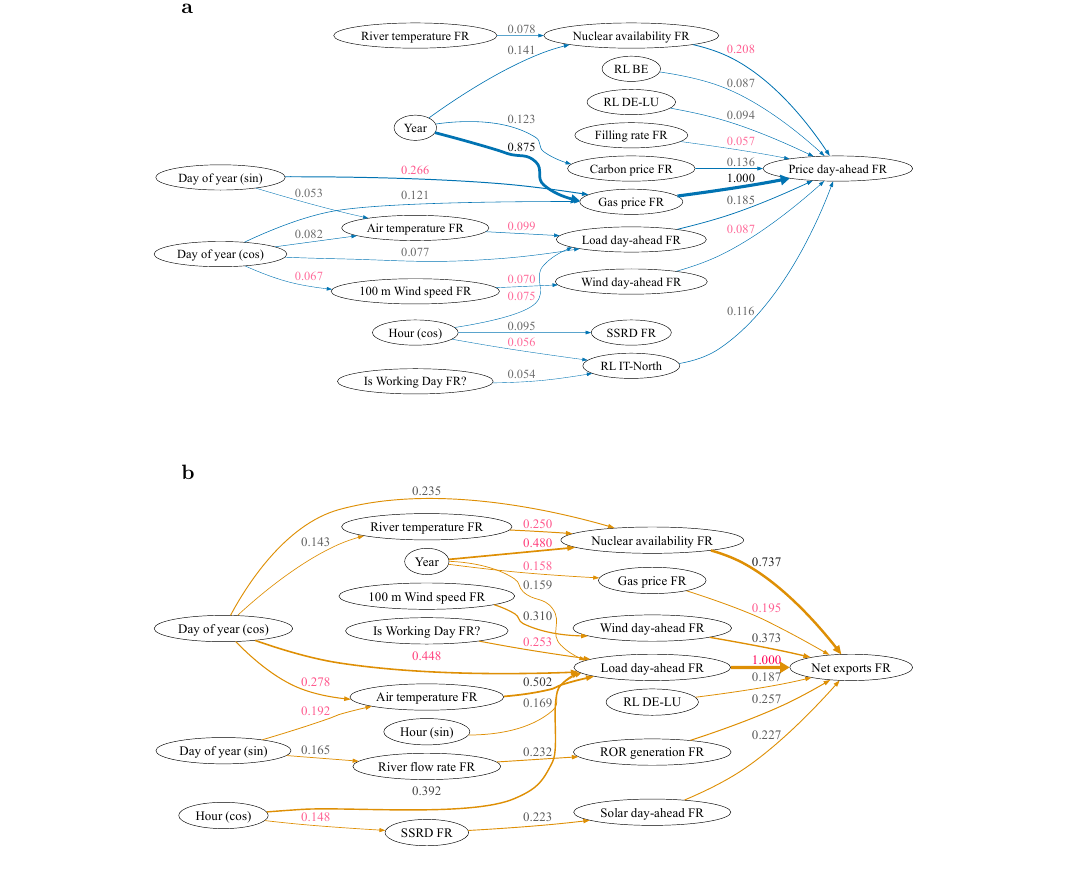}       
    \caption{
    \textbf{Mean absolute Shapley Flow values explain the non-linear GBT models for electricity market prices (\textbf{a}) and  net exports (\textbf{b}) of France.} 
    In contrast to the causal graph shown in Fig.~\ref{fig:structural causal model}, only the $25$ most important edges are shown for clarity. The edge attributions quantify how a given feature directly or indirectly influences the target prediction and is given in  \si{EUR/MWh} for the price and in \si{MW} for the net exports. The color indicates a positive (grey) or negative (red) correlation coefficient between Shapley Flow values on the respective edge and the feature value of the starting node.
    }
    \label{fig:Shapley_flows_main_example}
\end{figure*}

Linear SCMs are inherently transparent but they cannot account for non-linear relationships in the data. Therefore, we complement our analysis with a Gradient Boosted Tree (GBT) model, which can account for arbitrary, potentially complex, relationships. For several intermediate variables, this results in a marked improvement in predictive performance compared to the linear SCM, indicating pronounced non-linear behaviour (Supplementary Fig.~\ref{fig:r2-scores_gbt}).
While GBTs are not inherently transparent, we obtain efficient ex-post explanations of the model using SHapley Additive exPlanations (SHAP)~\cite{lundberg2020local}. Since standard SHAP values do not include causal knowledge, we utilize the Shapley Flow extension~\cite{wang2021shapley}, which explains the model in terms of a causal graph (see Methods).

Combining a causal graph with a GBT model provides further insights into the causal flow (Fig.~\ref{fig:Shapley_flows_main_example}). 
We focus again on France, while results for Spain are provided in Supplementary Fig.~\ref{fig:causal_graph_flow_spain}. Shapley Flow values are interpreted from right (target and leaf of the causal graph) to left (roots of the causal graph). For the electricity market price, we observe that the gas price has the strongest direct effect, with load and nuclear availability ranking second and third. For the net exports, day-ahead load and nuclear availability have the largest direct effect on the target. This is consistent with our domain understanding and the linear SCMs (Fig.~\ref{fig:structural_coefficients}\textbf{a,b}).

Moving to the left, and thereby closer to the root nodes, we note how the load itself is influenced by calendrical features and the air temperature feature in both, the price and the export models. Electric heating plays a major role in France contributing to a temperature-dependence of the load~\cite{heinen2018heat}. The other main directly relevant feature, the nuclear availability, is mostly affected by calendrical features and the river temperature, again in both models. The dependence on calendrical features accounts for seasonal effects and the decrease during the energy crisis. The dependence on river temperature is consistent with our domain understanding: High river temperatures limit the capability of thermal power plants to cool and hence generate electricity. 
This consistency with physical and operational knowledge provides additional confidence in the model and supports its use for the what-if analyses presented in the next section.

\subsection*{Indirect causal effects of temperature and river flow}

\begin{figure*}[ht]
    \centering
    \includegraphics[width=\linewidth]{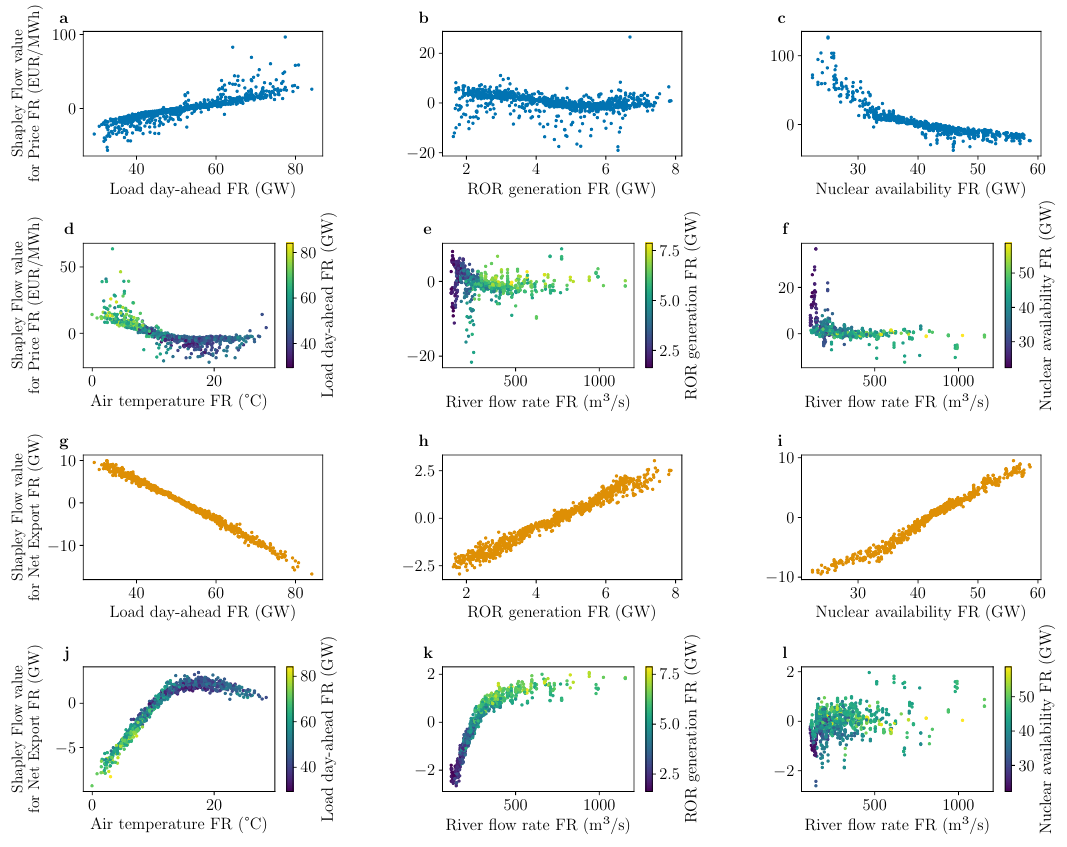}       
    \caption{
    \textbf{Shapley Flow analysis reveals direct and indirect non-linear dependencies.}     
    \textbf{a-c,} Shapley Flow dependencies plots for direct effects in the model for the French electricity market price. 
    \textbf{d-f,} Shapley Flow dependencies plots for indirect effects on the electricity market price. 
    Air temperature via Load, River flow rate via ROR generation and River flow rate via Nuclear availability. The value of the intermediate variable is color coded. 
    \textbf{g-i,} Shapley Flow dependencies plots for direct effect in the model for the French net exports. 
    \textbf{j-l,} Shapley Flow dependencies plots for indirect effects on the net exports. 
    Indirect effects are strongly non-linear.
    }
    \label{fig:dependence_plot}
\end{figure*}

Shapley Flow values explain the non-linear dependencies learned by the model and allow us to disentangle direct and indirect effects as shown in Fig.~\ref{fig:dependence_plot}. Based on the insights of the flows through the causal graph (Fig.~\ref{fig:Shapley_flows_main_example}), we specifically follow the flows from {air temperature} via the load, and the flows from river flow rate via run-of-river hydro generation and nuclear availability. We emphasise that also the indirect effects are still measured towards the target variable. 

The direct effect of load on both targets is approximately linear, whereas the indirect effect of air temperature via load is strongly non-linear (Fig.~\ref{fig:dependence_plot}\textbf{d,j}). At low temperatures, electricity market prices decrease and exports increase as air temperature rises. This reflects the role of electric heating in France, where electricity demand declines with increasing temperature. The dependence saturates for temperatures above \qty{10}{\degreeCelsius}, when heating demand becomes negligible. For temperatures exceeding \qty{20}{\degreeCelsius}, the relationship reverses, likely due to additional electricity demand from air conditioning.

River flow affects electricity markets through multiple channels, most prominently ROR hydropower generation and the availability of cooling water for nuclear plants. In recent years, French power plants have repeatedly faced operating restrictions during hot and dry summers due to insufficient cooling water~\cite{ahmad2021increase}. Climate projections suggest that precipitation and river flows in France may decline under continued high greenhouse gas emissions~\cite{van2013global}. Model-based studies have quantified the potential impact on thermoelectric and hydropower generation \cite{van2016power, PECHAN201463}. Here we provide an empirical analysis of the impact of river flows on electricity market prices.

The direct effects of ROR generation and nuclear availability on electricity market prices and net exports are weakly non-linear. In contrast, the indirect effects of river flow exhibit pronounced non-linearities. Along the pathway via ROR generation, prices decrease and exports increase as river flow rises, particularly below \qty{500}{m^3/s} (Fig.~\ref{fig:dependence_plot}\textbf{e,k}). This reflects a regime of water scarcity, where hydropower output is constrained by limited water availability. The effect saturates once river flow becomes abundant.

Non-linear effects are even more pronounced along the pathway via nuclear availability, but are concentrated at very low river flow levels (Fig.~\ref{fig:dependence_plot}\textbf{f,l}). In these rare situations, nuclear generation is curtailed due to insufficient cooling water, leading to substantial effects on electricity market prices and net exports. Overall, the nuclear pathway is less frequent and largely confined to exceptional conditions, whereas the ROR pathway operates over a broader range of flow levels. Nevertheless, both mechanisms are of comparable magnitude and should both be considered when assessing the impact of climatic conditions on electricity markets.
We present a sub-graph of the entire causal graph providing additional details on the effects of the meteorological variables in Supplementary Fig.~\ref{fig:zoom_in_causal_graphs}.

Summarising, GBT models with Shapley Flow provide additional insights and consistency checks. We find that the relations of the target and its parents in the causal graph are mostly linear, confirming the insights from the linear SCM. In contrast, Shapley Flow values reveal strongly non-linear indirect effects, as demonstrated for the temperature and the river flow rate. Notably, this analysis is not possible with linear SCMs or ordinary SHAP values~\cite{wang2021shapley}.

\subsection*{What-if scenarios}

\begin{figure*}
    \centering
    \includegraphics[width=\linewidth]{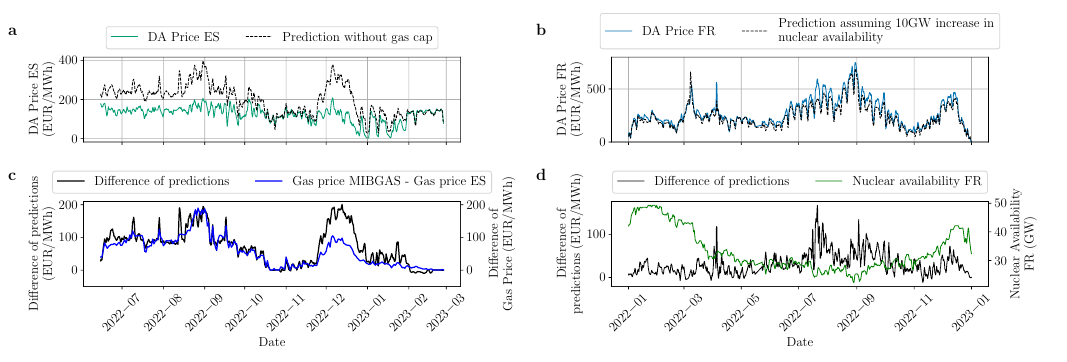}       
    \caption{
    \textbf{What-if Scenarios for day-ahead (DA) electricity market prices during the European energy crisis.}
    \textbf{a,} Actual electricity market prices in Spain are compared to model predictions for a hypothetical scenario where Spain did not establish a gas price cap.
    \textbf{c,} Difference in model predictions of the Spanish electricity market price (black line) using either the actual capped gas prices or the hypothetical uncapped gas price. This is compared to the uncapped gas price (blue line).
    \textbf{b,} Actual electricity market prices in France are compared to model predictions for a hypothetical scenario where nuclear availability was \qty{10}{GW} higher. 
    \textbf{d,} Difference in model predictions of the French electricity market price (black line) using either the actual nuclear availability or the hypothetical higher availability. This is compared to the actual nuclear availability.
    In all panels we plot daily averages for the sake of clarity.
    }
    \label{fig:what_if_scenarios}
\end{figure*}
Next, we turn from analysis of data and models to predictions. Our previous analysis of the GBT model with Shapley Flow values confirmed that it not only displays good model accuracy but also learned relationships that are consistent with our domain understanding. We use this validated model to consider hypothetical interventions during the energy crisis. 
Two historic facts inspire our research: First, the Spanish government introduced a gas price cap in June 2022 to limit the burden of high energy costs. Second, the availability of French nuclear reactors was at historically low values in 2022 as discussed in detail above.

What would have happened if the Spanish government had not enacted a gas price cap, as they did from June 2022 to February 2023~\cite{louIberianExceptionWhat2025, hidalgo-perezIberianExceptionEstimating2024}?
To address this question, we let the GBT model predict the electricity market price in Spain while assuming that there was no gas price cap. That is, we replace the actual Spanish gas prices by uncapped market prices in the model input for the period from 15 June, 2022, to 28 February, 2023. 
In this hypothetical scenario, we observe a strong increase in the electricity market price prediction in comparison to the actual price (Fig.~\ref{fig:what_if_scenarios}\textbf{a}). 
We further compare model predictions with actual gas prices and hypothetical uncapped gas prices in the input in Fig.~\ref{fig:what_if_scenarios}\textbf{c}. For the time period of approximately eight months, we observe that the electricity market price predictions average to $\SI{123.92}{EUR/MWh}$ and $\SI{195.62}{EUR/MWh}$, respectively. 
That is, the gas price cap corresponds to a reduction of average electricity market prices of $\SI{71.70}{EUR/MWh}$ or approximately $36.7\%$.
These values closely match estimates from a mechanistic market model, where the hypothetical scenario was included by distorting the actual supply curves, yielding a reduction by $\SI{67.61}{EUR/MWh}$~\cite{louIberianExceptionWhat2025}.
Finally, we compare the differences in model prediction to the difference in the model input in Fig.~\ref{fig:what_if_scenarios}\textbf{c}. We observe a very strong agreement until December 2022, and a clear but weaker correlation afterwards. The decrease in correlations in the second period may be attributed to increased nuclear generation in France and unusually warm weather~\cite{louIberianExceptionWhat2025}. 
We conclude that the Spanish government's intervention was successful in limiting electricity market price peaks. 

What would have happened if the French nuclear fleet had not suffered exceptionally low availability in 2022? To address this question, we consider a hypothetical scenario where nuclear availability was higher by \qty{10}{GW}, inspired by 10 reactors undergoing maintenance~\cite{Gaulier_Serfaty_2023}. That is, we increase the actual nuclear unavailability by \qty{10}{GW} in the model input for the period from 1 January, 2022, to 31 December, 2022. In this hypothetical scenario, we observe consistently lower electricity market price predictions in comparison to the actual price (Fig.~\ref{fig:what_if_scenarios}\textbf{b}), but extreme price peaks persist.
We further compare model predictions with actual nuclear availability and hypothetical increased nuclear availability as inputs in Fig.~\ref{fig:what_if_scenarios}\textbf{d}. The electricity market price predictions average to $\SI{278.28}{EUR/MWh}$ and $\SI{247.10}{EUR/MWh}$, respectively, with a difference of $\SI{31.18}{EUR/MWh}$. The difference peaks between July and September, when the nuclear availability was temporarily below $\qty{30}{GW}$, see Fig.~\ref{fig:what_if_scenarios}\textbf{d}. In this period, differences reach maximum values up to $\SI{278.13}{EUR/MWh}$ for a single hour and up to $\SI{167.27}{EUR/MWh}$ for the daily average.

We conclude that the impact of additional nuclear availability would have varied substantially over the course of 2022. The effect would have been strongest during the summer, when the French nuclear fleet reached historically low availability levels. This finding is consistent with our earlier analysis. In our dependency analysis, we record the largest Shapley flow values when nuclear availability drops below \SI{35}{GW}, see Fig.~\ref{fig:dependence_plot}. Hence, an increase in nuclear availability would have had the largest impact during periods when it would have been sufficient to fully replace gas-fired generation and imports. However, during many hours of operation, gas-fired plants and imports continued to play a decisive role even if additional nuclear generation would have been available. Consequently, averaged over the year 2022, an additional \SI{10}{GW} of nuclear availability would have reduced electricity prices by approximately \SI{31.18}{EUR/MWh}, or $11.2\%$, reflecting a limited aggregated effect.

Finally, we note that we used the GBT model to consider certain non-observed feature combinations, such as a high nuclear availability in summer. Such a procedure risks that the model has to extrapolate and hence display increased prediction errors. However, we have sufficient training data with similar values in nuclear availability and gas price, respectively, see Supplementary Fig.~\ref{fig:what-if_shapley_flow}. Hence, our validated GBT model can be used to consider hypothetical interventional scenarios to obtain quantitative insights.

\section*{Discussion}

To inform political decisions, transparency and empirical verification are essential in energy systems research~\cite{pfenninger2017energy}. 
In this article, we have demonstrated the importance and perspectives of causal statistical models in energy systems analysis. Simpson's Paradox (Fig.~\ref{fig:simpsons paradox}) shows that simple correlation studies can be highly misleading and thus inadequate for empirical studies. Causal graphs resolve this paradox and pave the way for causal machine learning models. We have applied such models to the French electricity market, which was severely affected during the European energy crisis. 

We emphasise that the models of French and Spanish electricity market prices and French net exports show good predictive performance for their respective final targets, while the performance for intermediate nodes varies between variables.
We remark that this is not crucial in the current setting because we mainly focus on the results for electricity market price and net exports.

We identify two main factors underlying the sharp rise in French electricity market prices: reduced nuclear availability and the surge in natural gas prices. With up to half of the nuclear fleet unavailable, electricity imports increased substantially and gas-fired generation played a larger role in meeting domestic demand. As a consequence,  electricity market prices in France and its neighbouring countries often synchronised and gas-fired plants played a decisive role in setting the market clearing price. French electricity market prices were therefore highly exposed to elevated gas prices, similar to its neighbours. Both the SCM and the GBT analysis consistently identify these mechanisms and enable a quantitative assessment of their respective contributions.

Our analysis further highlights the importance of cross-border interactions in interconnected electricity markets. We show that developments in the French nuclear fleet propagated to Spain and that cross-border residual load effects contribute measurably to price formation. 

Finally, the causal framework allows us to explore selected what-if scenarios in a structured manner. By modifying selected inputs in the validated GBT model, we examined how electricity market prices would have evolved in the absence of the Spanish gas price cap and under higher nuclear availability in France. In this way, we estimate that the Spanish gas price cap reduced electricity market prices by an average of \SI{71.10}{EUR/MWh}, in close agreement with estimates from mechanistic market models reported in the literature. The effect of increased nuclear availability in France would have varied markedly over time. During certain periods in the summer of 2022, an additional \SI{10}{GW} would have been sufficient to replace gas-fired generation and imports, leading to pronounced price reductions. However, over most of 2022 the impact would have remained limited.

Beyond our contributions to understanding the effects of the energy crisis, we further advocate the usage of structural causal models and causal explanations of machine learning models. These tools provide causal insights into complex problems, as we often encounter them in the energy system.  Moving forward, our methods, both linear SCMs and non-linear GBTs, could be applied to other countries, assuming sufficient data is available.

\section*{Methods}
\footnotesize

\subsection*{Data sources and pre-processing}

Time series data on day-ahead wind generation, day-ahead solar generation, day-ahead load, actual run-of-river and poundage (ROR) generation, installed nuclear capacity, filling rate of hydro storage, scheduled commercial exchange and day-ahead electricity price were obtained from the ENTSO-E transparency platform \cite{entsoeElectricityMarket}. The data was retrieved using the ENTSO-E restful API using the open source Python package entsoe-py \cite{githubGitHubEnergieIDentsoepy} as well as partly manually downloaded. 
We included the residual load of neighbouring countries who participate in the Single Day-Ahead Coupling (SDAC) of electricity markets throughout the period of our study~\cite{entso-e_single}.
Multiple bidding zone changes occurred in Italy and concerning the bidding zones DE-AT-LU and DE-LU, which were considered as described in the Supplementary Methods. 
The aggregated net exports were calculated using the total scheduled exchange considering all neighbours. 
In addition, Spain has the non-SDAC neighbours Morocco and Andorra. Data of these exchanges can be found using the API of Red Électrica \cite{REE_API}.
RTE \cite{RTE_IIP_production_unavailability} provides data on nuclear unavailability in case of France and was pre-processed as described in the Supplement. The nuclear availability is obtained by subtracting the unavailable nuclear capacity from the total installed nuclear capacity provided by ENTSO-E. In case of Spain, ESIOS of Red Électrica \cite{REE_ESIOS_analysis} provides data on nuclear availability directly.

In this work, the European emission trading system (EU ETS) \cite{ec_ets_price_transmission_2025}  is referred to as carbon price. The World Group Bank \cite{worldbankPrice} provides EU ETS in $\si{US\$/tCO_2e}$ which was converted to $\si{EUR/tCO_2e}$ at a fixed exchange rate of 0.93 $\si{EUR/US\$}$ on 7 May 2024~\cite{finanzenDollarEuro}.

Data on the share of natural gas in the electricity mix and the dependency on Russian gas imports are available at Eurostat~\cite{europa}.
The gas price data for France was obtained from the TTF daily futures~\cite{thomsonreuters}. Missing daily gas prices were interpolated. Raw data was converted from $\si{US\$}$ to $\si{EUR}$ using the aforementioned exchange rate.
The gas price of Spain was calculated using the MIBGAS gas prices \cite{MIBGAS_file_access} and capping them according to the Iberian exception as described by the European Commission \cite{EU_Commission_IP_23_2408}.

Weather data including surface solar radiation downwards, wind speed at \qty{100}{\meter} and air temperature at \qty{2}{\meter} is provided by ERA5 \cite{Hersbach_ERA5_2023, hersbach2020era5}, mean values are calculated by averaging over larger areas including France and Spain respectively (Supplementary Fig.~\ref{fig:french grid}). 
The API Hub'Eau~\cite{eaufranceAccueilHubeau} provides hourly data of river temperatures and daily data of river flow rates in France. Mean river flow and river temperature were averaged, while only considering rivers near nuclear power plants (see Supplementary Methods for details). 
In case of Spain the river flow and temperatures of the river Ebro are gathered from SAIH Ebro \cite{CHE_SAIH_Ebro_historicos}.

The binary variable ``Is working day?'' of the linear causal models is True for working days. Data on French public holidays was gathered from Ref.~\cite{dataJoursFris}, and Spanish data from Ref. ~\cite{officeholidays_spain}. The cyclical variables
``day of year'' and ``hour of day'' are parameterised by sinus and cosine functions.
To be able to compare the periods before and during the European energy crisis, we must set a reference date for the beginning. Here we choose October 2021, the first month in which the average price in France was significantly higher ($>170 \, \si{EUR}/\si{MWh}$) than the average price of the previous year ($64 \, \si{EUR}/\si{MWh}$).

\subsection*{Implementation of SCMs}

We use the DoWhy Python package~\cite{dowhy, JMLR:v25:22-1258} to perform the causal analysis. First a causal model is created based on the causal graph shown in Fig.~\ref{fig:structural causal model}. The corresponding graph is implemented with the Networkx Python package~\cite{networkx}. 

Then, a structural causal model is created by assigning a causal mechanism to every node. Within the model, root nodes are assumed not to be caused by anything. They are thus assigned an empirical distribution sampled from the normalised data. To non-root nodes a linear regression model with additive noise is assigned, where the noise accounts for unobserved variables. The model is fitted to the normalised data. Model evaluation results are discussed in Supplementary Note~2. 

The data is normalised by subtracting the mean and dividing by the standard deviation of each variable to ensure numerical stability during fitting. The normalised data is thus given by
\begin{align}
    X_{j, \rm norm} = (X_j - \left\langle X_j \right\rangle)/\sigma_{X_j}.
\end{align}
The structural coefficients resulting from the fitting are then denormalised after fitting. This is achieved by multiplying by the standard deviation
\begin{align}
    c_{ij} = c_{ij,\rm norm}\cdot \frac{\sigma_{X_i}}{\sigma_{X_j}}.
\end{align}

\subsection*{GBT models}

For the GBT models, we use the XGBoost Python package~\cite{chenXGBoostScalableTree2016}. For each intermediate node as well as for the target nodes of the causal graph, a GBT model is trained which uses the parent node features as inputs. The GBT models for the targets are further used for the what-if scenarios.
In order to find the best hyperparameters for training the gradient boosting trees, we carry out a random search. Further, the models are trained on 80 percent of the data while the remaining 20 percent is used for testing. 
To prevent overfitting and data leakage, the time series data is split in the following way into test and train sets: all data points are grouped into one-day intervals, and each interval is randomly assigned to the test or train set according to the 80 percent split percentage. 

\subsection*{Shapley Flow}

SHapley Additive exPlanations (SHAP)~\cite{lundberg2020local} is one of the most widely used approaches for explaining machine learning models and is based on Shapley values from cooperative game theory~\cite{shapley1953value}, which distribute the total gain of a game fairly among players according to their individual contributions.. SHAP values attribute a model prediction to the input features by quantifying their marginal contributions across all possible feature coalitions. These attributions satisfy several desirable axiomatic properties, including efficiency, linearity, and the null-player property. However, standard SHAP treats features as independent players and therefore does not explicitly account for causal dependencies between variables.

SHAP values $\phi_i(x)$ of the feature $i$ are calculated as follows:
\begin{equation*}
\phi_i(x) = \sum_{S \subseteq V \setminus \{i\}}
\frac{|S|! \,(|V| - |S| - 1)!}{|V|!}
\left[ f\big(x_{S \cup \{i\}}\big) - f(x_S) \right]
\end{equation*}
where $f$ denotes the model output function, $V$ the set of all input features, and $x$ an input instance. Furthermore, $f(x_S)$ represents the expected model output conditioned on the features contained in subset $S$.

Shapley Flow extends the SHAP framework to settings where dependencies between features are explicitly modeled via a directed causal graph~\cite{wang2021shapley}. Instead of attributing the prediction directly to features, Shapley Flow assigns attribution values to the edges of the causal graph, thus quantifying how predictive information propagates through causal pathways. This allows the explanation to distinguish between direct and indirect contributions and provides a more causal interpretation of the model output.

A key concept of Shapley Flow is the explanation boundary, which partitions the causal graph into an upstream and a downstream region. 
For a given boundary, the model output is evaluated under a partial activation of edges, where upstream variables take foreground values and downstream variables take background values. The resulting change in output quantifies the contribution associated with that boundary. By averaging these contributions over all possible boundary configurations according to the Shapley principle, attribution values are obtained for individual edges in the graph.

Shapley Flow assigns an attribution $\phi_e$ to each edge $e \in E$ of the causal graph $G=(V,E)$ by estimating its expected marginal contribution over both graph traversal configurations and background samples. Given a foreground input $x$ and a set of $n_b$ background samples $\{x_0^{(j)}\}_{j=1}^{n_b}$, the attribution is computed as a nested expectation, where the inner expectation averages the marginal contribution $\Delta_e(\pi, x_0^{(j)})$ over $n_{\mathrm{runs}}$ randomly sampled graph traversal configurations $\pi$, and the outer expectation averages over the background samples. In practice, this is approximated via Monte Carlo sampling as 
\begin{equation}
    \phi_e = \frac{1}{n_b n_{\mathrm{runs}}}\sum_{j=1}^{n_b}\sum_{i=1}^{n_{\mathrm{runs}}}\Delta_e(\pi_i, x_0^{(j)}),
\end{equation}
where $\Delta_e(\pi, x_0)$ denotes the difference in the model output when edge $e$ is added to the set of preceding edges in configuration $\pi$, evaluated with respect to foreground input $x$ and background input $x_0$.
In our experiments, we use $1000$ foreground samples, $n_b=96$ background samples, and $n_{\mathrm{runs}}=750$ graph configuration samples.

A graph configuration is an ordering of each node’s outgoing edges for depth‑first search (DFS). For each ordering, DFS runs from the source and processes edges in that order, updating target nodes with values from their sources. When the sink is reached, the change in its output is attributed to all edges along the path. The final result is the average of these attributions over all configurations.

Shapley Flow preserves fundamental axiomatic properties of Shapley values, such as efficiency, linearity, and the null-player property. In addition, it satisfies a boundary consistency property, which ensures that edge attributions remain consistent across different boundary configurations. As a result, the model prediction can be decomposed into a set of causal information flows that reveal the relative importance of different pathways in the underlying causal graph. 
We implement Shapley Flow following Wang et al.~\cite{wang2021shapley}. Further implementation details are provided in our code repository \cite{zenodo_causal_repo2025}.

\subsection*{AI declaration.} 
Generative AI tools, specifically ChatGPT and GitHub Copilot, were used to assist with language editing and portions of the code development. All outputs were reviewed and validated by the authors, who take full responsibility for the content.

\subsection*{Data availability.} 

Natural gas prices with daily resolution were obtained from Ref.~\cite{thomsonreuters} and are subject to copyright. All other data used in this study are publicly available from the respective sources as described in the Methods section. Detailed scripts to gather relevant data are also available on Zenodo~\cite{zenodo_causal_repo2025}.

\subsection*{Code availability.} 

Computer code is available on Zenodo~\cite{zenodo_causal_repo2025}.

\subsection*{Acknowledgements}
We thank Max Kleinebrahm and Wolf Fichtner for stimulating discussions.

\subsection*{Funding}

We gratefully acknowledge funding from the Helmholtz Association and the Networking Fund through Helmholtz AI and under grant no.~VH-NG-1727. This work was partially performed on the computational resource bwUniCluster funded by the Ministry of Science, Research and the Arts Baden-Württemberg and the Universities of the State of Baden-Württemberg, Germany, within the framework program bwHPC. This work was partially performed as part of the Helmholtz School for Data Science in Life, Earth and Energy (HDS-LEE) and received funding from the Helmholtz Association of German Research Centres.

\subsection*{Author Contributions}

D.W.~conceived research. D.W. and B.S. designed and supervised research with contributions by A.P.
S.S. and A.T. carried out the initial statistical analysis, implemented and evaluated the linear SCM with initial contributions by J.T. 
F.I. and U.O. implemented the GBT model and evaluated the Shapley Flow values. 
S.S. and U.O. designed all figures with inputs by A.T. and F.I.
All authors contributed to discussing the results and writing the manuscript.´

\subsection*{Competing Interests} 
The authors declare no competing financial or non-financial interests.

\clearpage

\section*{Supplementary Information}

\renewcommand{\figurename}{\textbf{Supplementary Figure}}
\renewcommand{\tablename}{\textbf{Supplementary Table}}
\makeatletter
\def\l@subsubsection#1#2{}
\makeatother

\refstepcounter{notecounter}
\subsection*{Supplementary Note \thenotecounter:~Seasonality of nuclear availability, load and energy prices}

Essential variables in the French electricity system display a strong seasonality as shown in Supplementary Fig.~\ref{fig:na_seasonality}. The river flow shows strong seasonal patterns. The demand for electricity is usually higher in the winter due to the prevalence of electric heaters~\cite{heinen2018heat}. Electricity market prices are determined by the balance of supply and demand, such that the price generally increases with the demand. In fact, we observe that prices are usually higher in winter than in summer.

The French electricity market heavily relies on nuclear power plants that require revisions. The demand and price are typically lowest during the summer and so are opportunity costs for the unavailability of nuclear power plants. Indeed we observe a pronounced seasonality of nuclear availability which closely follows the seasonality of the load (Supplementary Fig.~\ref{fig:na_seasonality}). 

Seasonal affects can induce confounding. Typically, both the nuclear availability and the electricity market price are highest in winter as argued above. This can lead to a perceived violation of economic laws if one only considers the supply side and neglects the seasonal variation in demand.

\begin{figure}[b]
     \includegraphics[width=\linewidth]{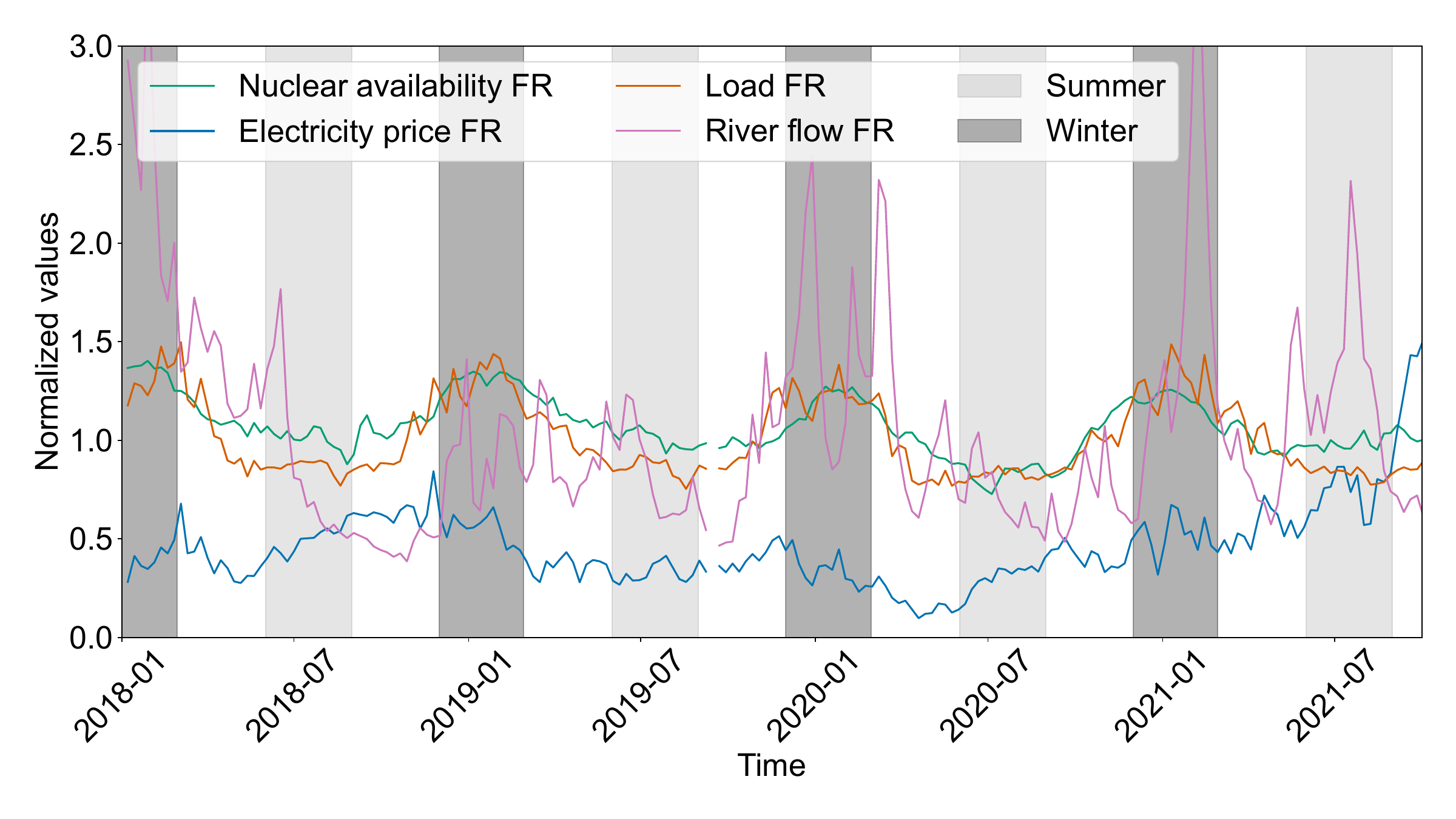}          
     \caption{\textbf{Seasonality in the French electricity system.} We plot the weekly averages of the nuclear availability, river flow, load day-ahead, and the day-ahead electricity market price before the start of the energy crisis normalised by their respective overall means. Winter and Summer are indicated with grey areas. River flow shows strong seasonal patterns. Nuclear availability is highest in the winter, which can be explained by revisions being scheduled in summer when load is lowest. Electricity market prices are higher in case of higher load, causing positive correlation between nuclear availability and electricity market prices.}
     \label{fig:na_seasonality}
\end{figure}

\refstepcounter{notecounter}
\subsection*{Supplementary Note \thenotecounter:~Evaluation of SCMs}

\begin{table}[tb]
    \centering
    \begin{tabular}{|c|c|c|c|c|}
        \hline
        Period & R\textsuperscript{2} & MAE & RMSE & LMC  \\ 
        & & & & Violations
        \\ \hline 
        \multicolumn{5}{|c|}{\textbf{Target: Price day-ahead FR (EUR)}} \\ \hline
        2018-01-01 – 2023-12-31 & 0.92 & 20.32 & 30.98 & 219/352 \\ \hline
        \multicolumn{5}{|c|}{\textbf{Target: Net exports FR (MW)}} \\ \hline
        2018-01-01 – 2023-12-31 & 0.83 & 1668.69 & 2133.83 & 222/352 \\ \hline
        \multicolumn{5}{|c|}{\textbf{Target: Price day-ahead ES (EUR)}} \\ \hline
        2018-01-01 – 2023-12-31 & 0.91 & 13.63 & 18.46 & 171/298 \\ \hline
    \end{tabular}
    \caption{Model Performance of SCM's for both electricity market price and net exports as target.}
    \label{tab:model_performance}
\end{table}

\begin{table}[tb]
    \centering
    \begin{tabular}{|c|c|c|c|c|}
        \hline
        Period & $R^2$ & MAE & RMSE & Mean Label \\
        \hline
        \multicolumn{5}{|c|}{\textbf{Target: Price day-ahead FR (EUR)}} \\
        \hline
        2018-01-01 -- 2023-12-31 & 0.96 & 10.71 & 22.95 & 96.21 \\
        \hline
        \multicolumn{5}{|c|}{\textbf{Target: Net exports FR (MW)}} \\
        \hline
        2018-01-01 -- 2023-12-31 & 0.92 & 1236.13 & 1579.03 & 4843.20 \\
        \hline
        \multicolumn{5}{|c|}{\textbf{Target: Price day-ahead ES (EUR)}} \\
        \hline
        2018-01-01 -- 2023-12-31 & 0.95 & 8.03 & 14.13 & 79.01 \\
        \hline
    \end{tabular}
    \caption{GBT performance metrics 
    for both electricity market price and net exports as target.
    }
    \label{tab:model_performance_gbt}
\end{table}

DoWhy~\cite{dowhy, JMLR:v25:22-1258} checks how many local Markov conditions (LMC) are violated in a given graph using conditional independence (CI) tests. During CI tests, the conditional independence of all non-descendants of each node given its parents is checked~\cite{eulig2024falsifyingcausalgraphsusing}.

Using LMCs, wrong graphs can be rejected, but a graph can not be declared as true with certainty. A benchmark is required of how many LMC violations are acceptable, above which graphs are rejected. To get a benchmark, we randomly permute the nodes of a given graph and test for violations of LMCs. The p-value, denoted as $p_{LMC}$, is defined as the probability of a randomly permuted graph to have fewer LMC violations than the original graph. We accept a given graph if the p-value falls below some threshold; otherwise it is rejected. 
Graphs of the same Markov equivalence class (MEC) have the same conditional independencies~\cite{pearl2016causal}. Thus, if too many of the permuted graphs lie in the same MEC, the conditional independence tests are not informative of the graph's correctness. A second p-value $p_{MEC}$ is defined as the probability of a randomly permuted graph to lie in the same MEC as the given graph. The graph is considered falsifiable if the p-value is below a certain threshold.
We use a significance level of 0.05 for both p-values. 
In conclusion, we say a graph is falsifiable if no more than 5~\% of the randomly permuted graphs lie in the same MEC as the given graph. We say a graph is falsified if more than 5~\% of the randomly permuted graphs have as few or fewer LMC violations as the given graph.

For all models (electricity market prices of France and Spain and net exports of France), the p-value $p_{MEC}$ is zero for 50 random node permuted graphs. None of the randomly permuted graphs lie in the same MEC as the given graphs. Therefore, we conclude that the corresponding causal graphs are falsifiable. To falsify the models, the violation of LMCs are compared against 50 random node permuted graphs. Out of the 352 possible LMC violations in case of the French model, the one for the electricity market prices violates 219 and the one for the net exports 222. The model for Spanish electricity market prices violates 171 out of 298 possible LMC violations. Nevertheless, the p-value $p_{LMC}$ is zero for all models. Therefore, none of the permuted graphs violate as few or fewer LMCs than the original graph. The causal graph thus performs significantly better than random and we accept it.

\begin{figure*}[tb]
    \centering
    \begin{minipage}{0.32\textwidth} 
        \begin{flushleft} 
            \textbf{a}
        \end{flushleft}
        \centering
            \includegraphics[width=\linewidth]{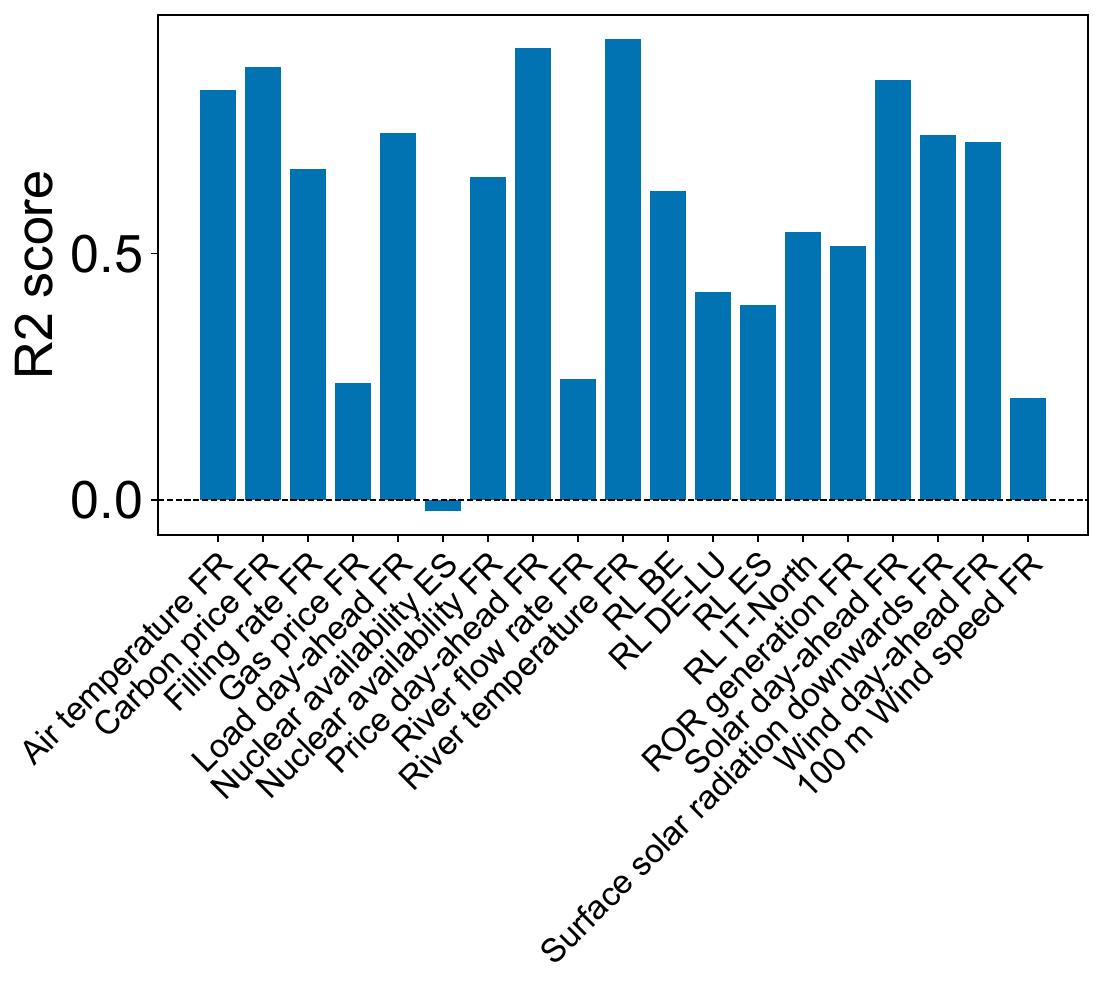}       
    \end{minipage}
    \begin{minipage}{0.32\textwidth} 
        \begin{flushleft} 
            \textbf{b}
        \end{flushleft}
        \centering
            \includegraphics[width=\linewidth]{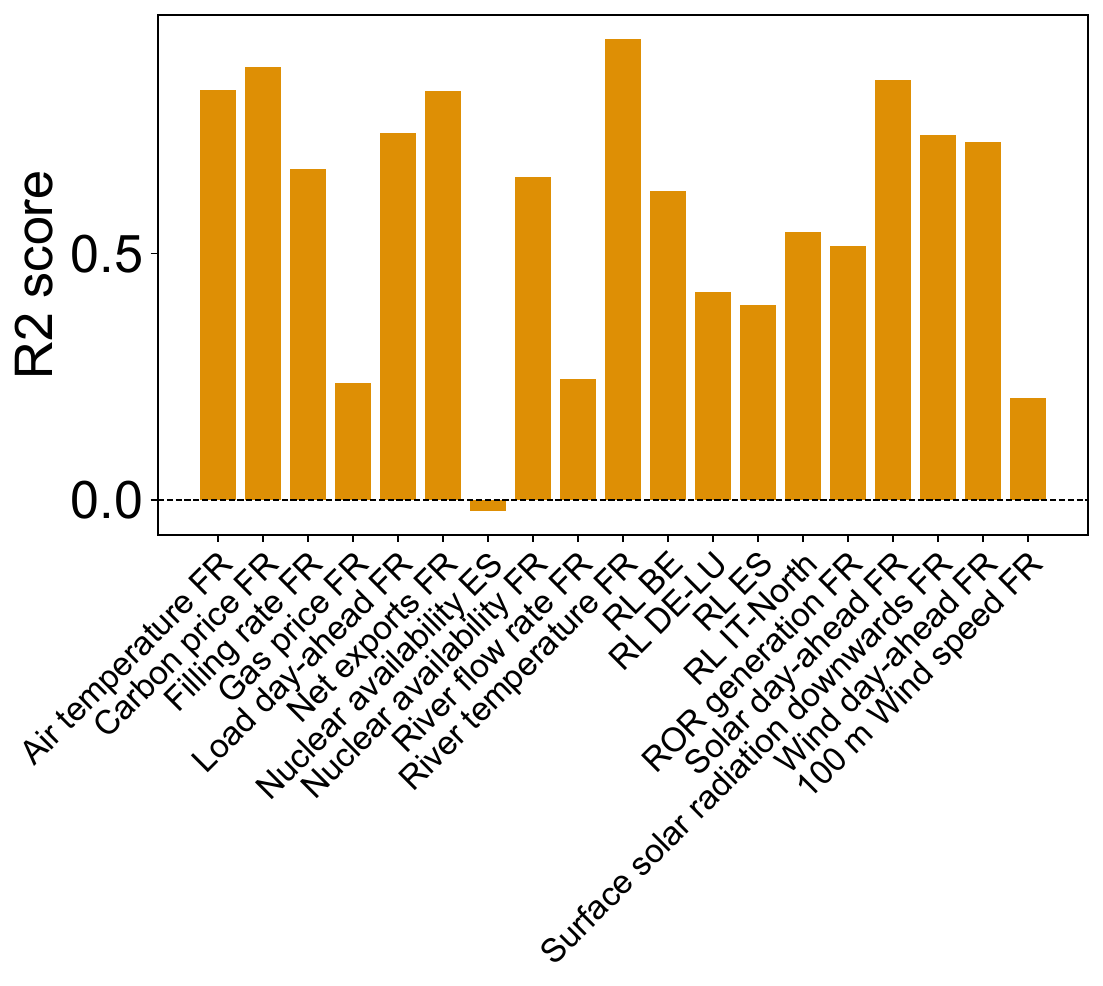}   
    \end{minipage}
    \begin{minipage}{0.32\textwidth} 
        \begin{flushleft} 
            \textbf{c}
        \end{flushleft}
        \centering
            \includegraphics[width=\linewidth]{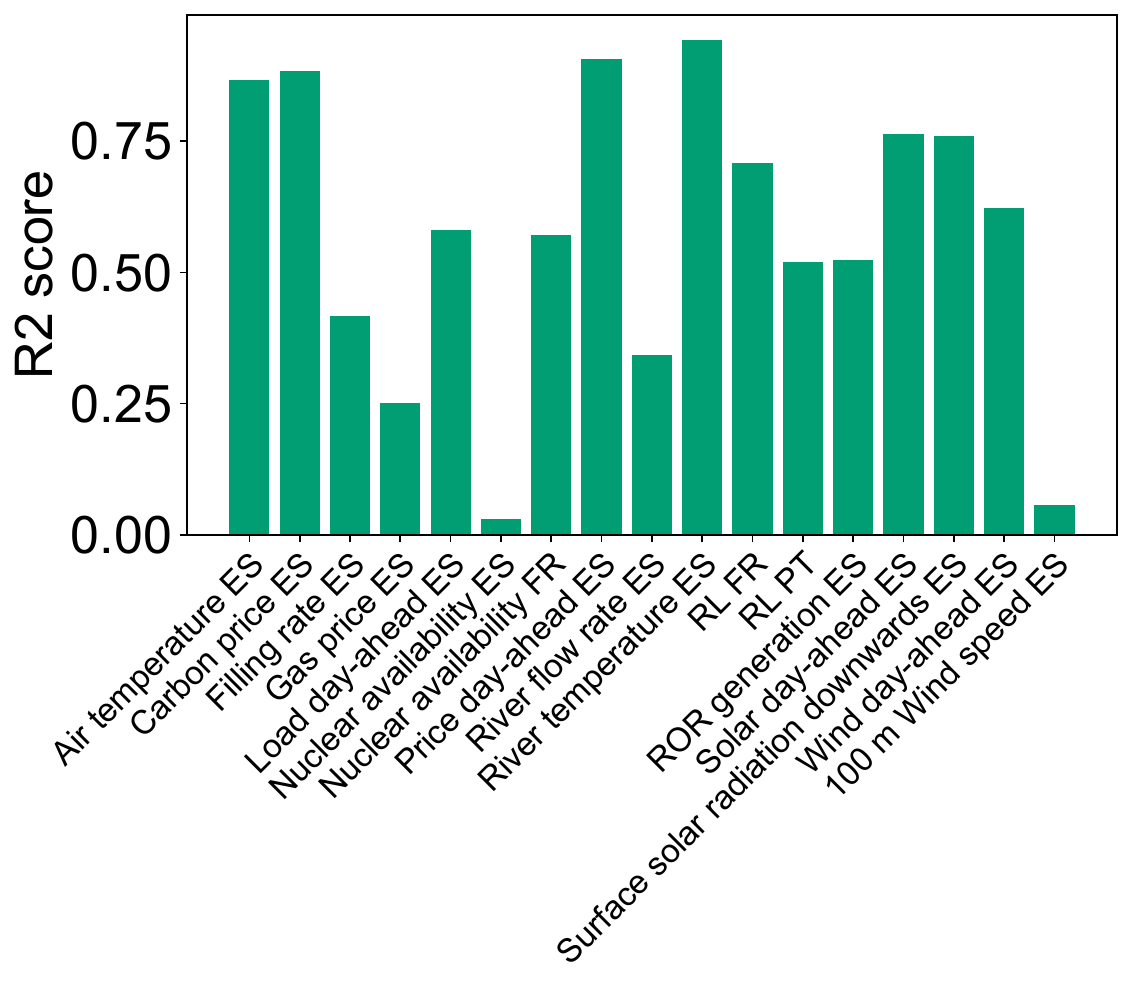}   
    \end{minipage}
    \caption{
        \textbf{$R^2$ scores of non-root nodes of SCMs.}
    $R^2$ scores of SCM with electricity market price of France \textbf{a}, net exports of France \textbf{b} and electricity market price of Spain \textbf{c} as target. The $R^2$ score of 0.92 of the French electricity market price and 0.83 of the french net exports and 0.91 of the Spanish electricity market price indicate the suitability of the causal graphs as well as the choice of a linear models.
        }
    \label{fig:R2 scores}
\end{figure*}

\begin{figure*}[tb]
    \centering
    \begin{minipage}{0.32\textwidth} 
        \begin{flushleft} 
            \textbf{a}
        \end{flushleft}
        \centering
            \includegraphics[width=\linewidth]{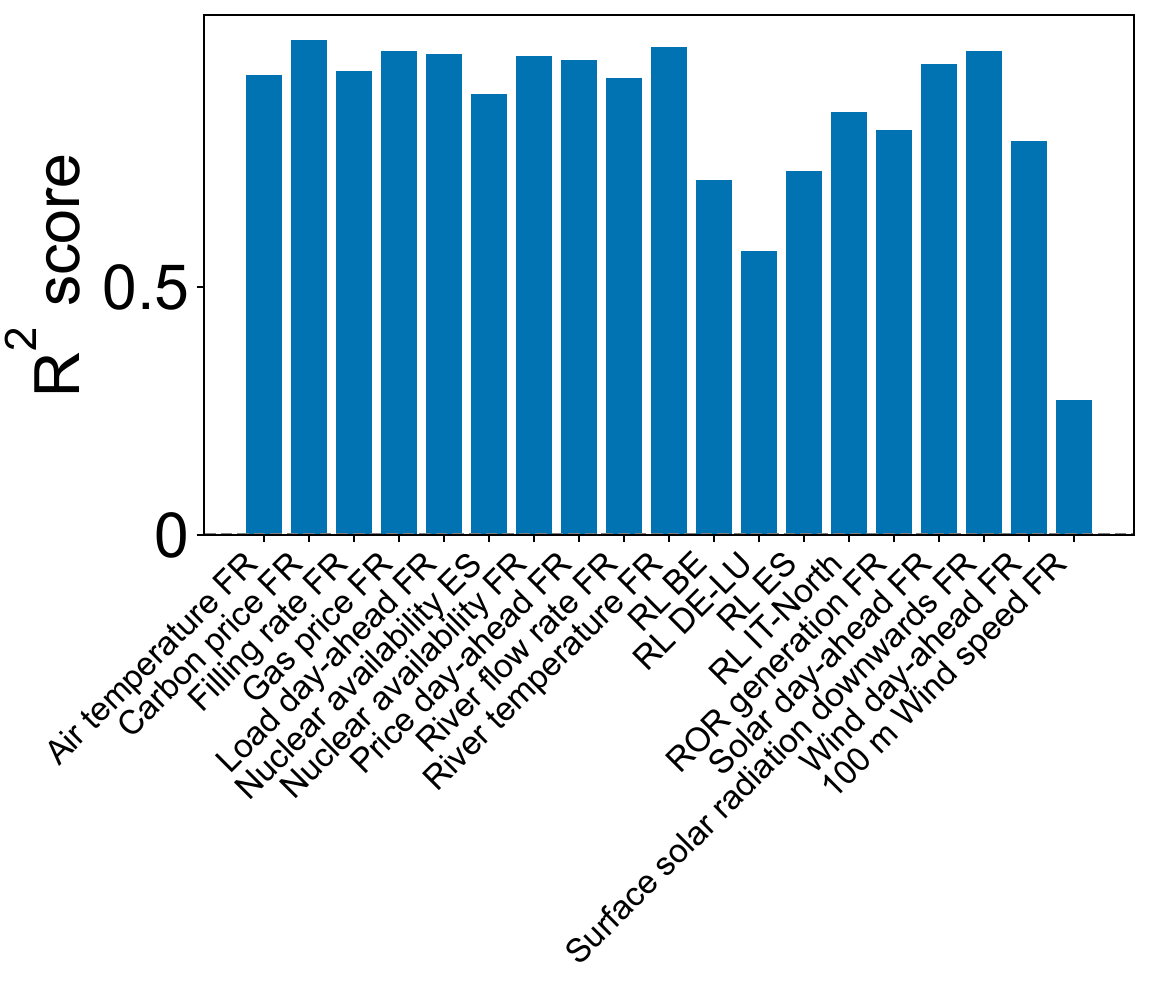}       
    \end{minipage}
    \begin{minipage}{0.32\textwidth} 
        \begin{flushleft} 
            \textbf{b}
        \end{flushleft}
        \centering
            \includegraphics[width=\linewidth]{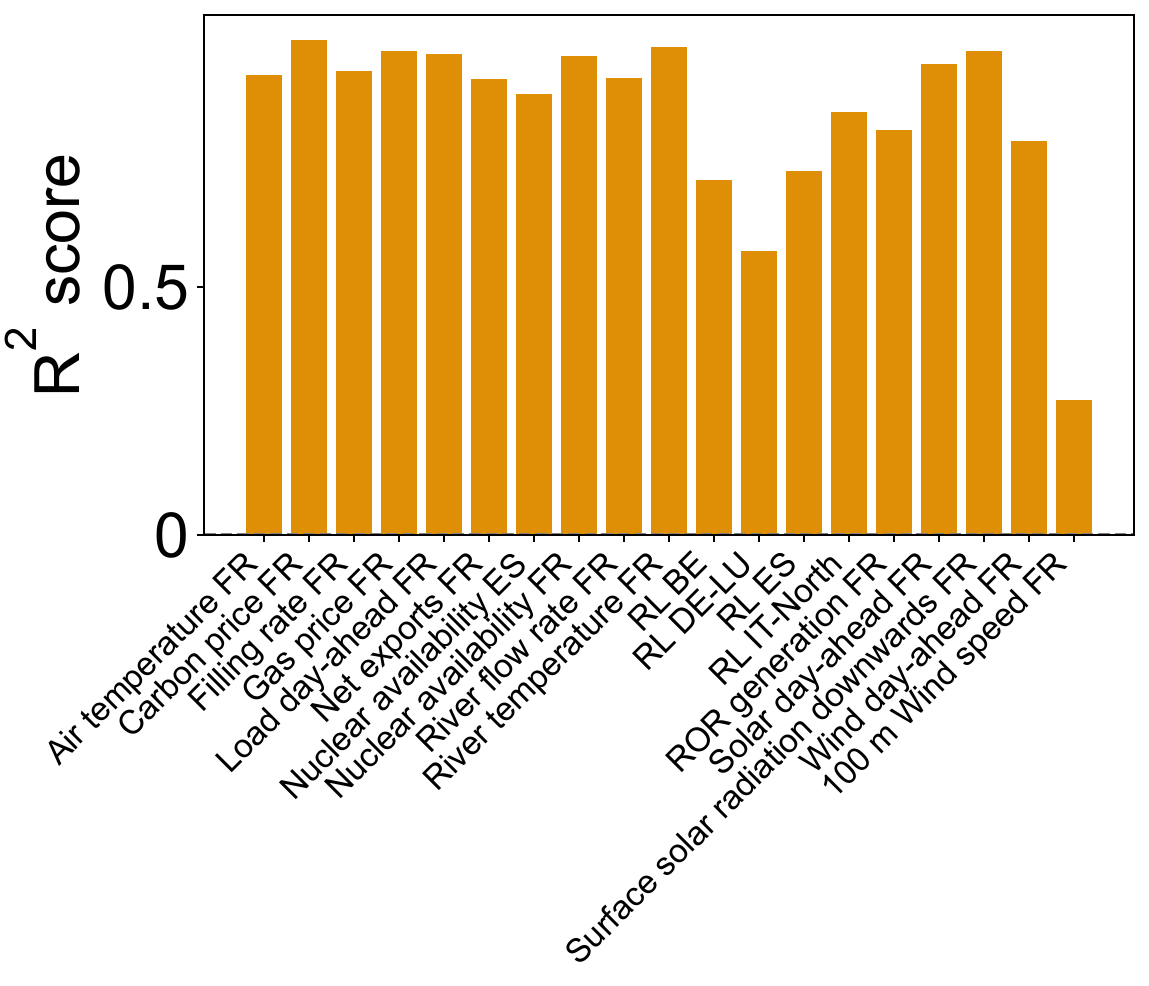}   
    \end{minipage}
    \begin{minipage}{0.32\textwidth} 
        \begin{flushleft} 
            \textbf{c}
        \end{flushleft}
        \centering
            \includegraphics[width=\linewidth]{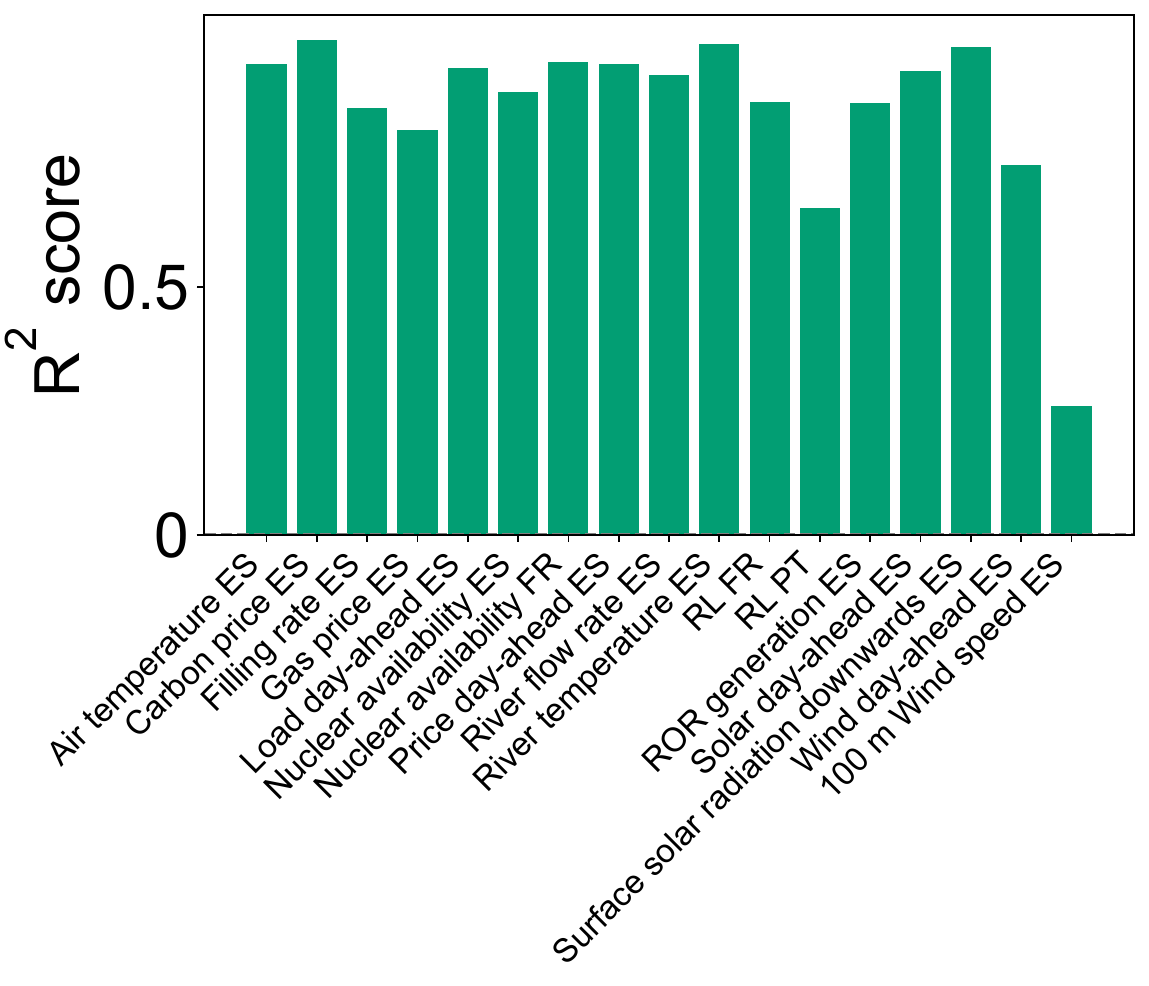}   
    \end{minipage}
    \caption{\textbf{$R^2$ scores for the GBT models of non-root nodes.}
    $R^2$ scores of GBTs with electricity market price of France \textbf{a}, net exports of France \textbf{b} and electricity market price of Spain \textbf{c} as target. The high $R^2$ scores of the non-root nodes demonstrate the successful explanation by the non-linear GBT models.
    }
    \label{fig:r2-scores_gbt}
\end{figure*}

We use the $R^2$ score to evaluate goodness of the fit for every structural equation. The $R^2$ score is defined as
\begin{align}
    R^2 = 1 - \frac{RSS}{TSS}.
\end{align}
Hereby, the total sum of squares $TSS$ is the squared difference between observed values and their mean. The residual sum of squares $RSS$ is the squared difference between model prediction and observed values. The ratio of $RSS$ and $TSS$ thus yields the fraction of variance that is not explained by the model. The $R^2$ score thus is the proportion of variance that the model can explain~\cite{draper1998applied}. To calculate the $R^2$ value the data is split into $80 \%$ training data and $20 \%$ test data. The split is performed in weeks to minimise over fitting.

Different effects can lead to low $R^2$ scores. The used linear model might not be appropriate in case of strongly non-linear causal effects. Further, relevant variables might be missing which are collectively represented in the noise term in the structural equations. This can lead to the noise term dominating in the structural equation. This mechanism is especially prevalent in highly complex systems featuring many variables and dependencies which can not all be considered in the model. As the electricity market is such a highly complex system, we need to decide which dependencies the model should include and for which variables we accept that they are mostly influenced by noise terms.

The $R^2$ scores of all fits are depicted in Supplementary Fig.~\ref{fig:R2 scores}. As our analysis focuses on the final targets, the electricity market prices and the net exports, their corresponding results are most important. We find a $R^2$ score of 0.92 for the electricity market price of France,  0.91 for the electricity market price of Spain and 0.83 for net exports of France. Hence, the proportion of variation in the target variables that the model can explain is high. The choice of a linear model, as well as the causal graph, appear to be a reasonable approximation. 

All variables relevant to assess model evaluation, the $R^2$ scores  and other metrics of the targets as well as the LMC violations, are listed in Supplementary Tab.~\ref{tab:model_performance}. The $R^2$ score of the nuclear availability of France is with 0.66 substantially lower than the $R^2$ scores of the target variables. This is to be expected since many influencing factors are not included in the model. For example, unforeseen corrosion damage lead to many reactors going into revision in 2021 and 2022~\cite{rtefranceFrenchAnnual}. These variables are not included in the model.

\refstepcounter{notecounter}
\subsection*{Supplementary Note \thenotecounter:~Structural coefficients of SCMs}

\begin{figure*}[tb]
    \centering
     \begin{minipage}{0.32\textwidth} 
        \begin{flushleft} 
            \textbf{a}
        \end{flushleft}
       \centering
            \includegraphics[width=\linewidth]{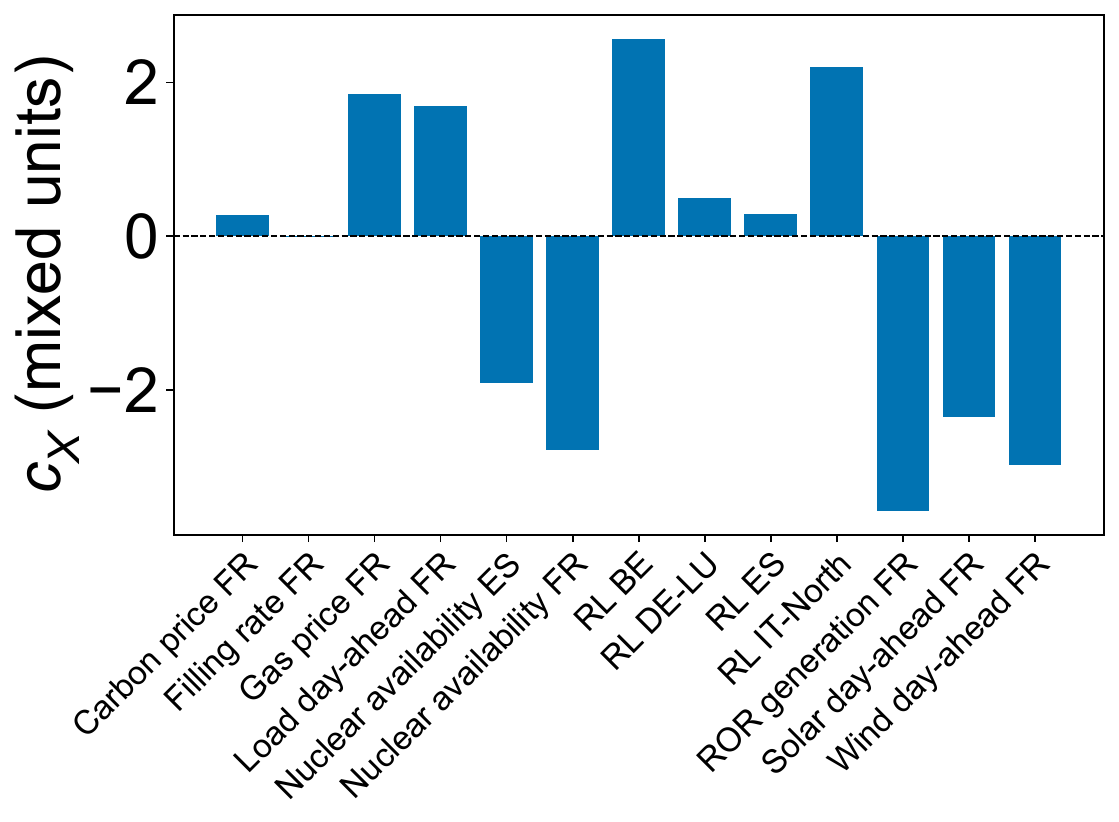}       
    \end{minipage}
    \begin{minipage}{0.32\textwidth} 
        \begin{flushleft} 
            \textbf{b}
        \end{flushleft}
        \centering
            \includegraphics[width=\linewidth]{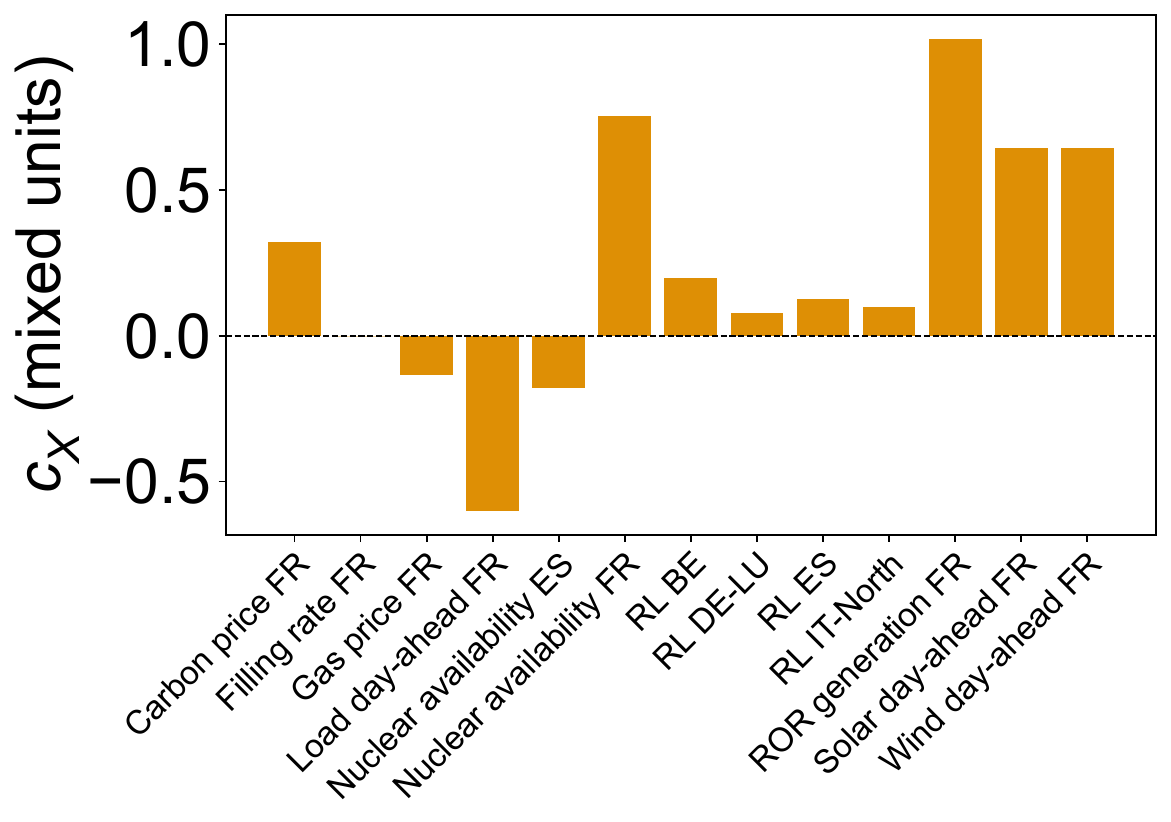}   
    \end{minipage}
    \begin{minipage}{0.32\textwidth} 
        \begin{flushleft} 
            \textbf{c}
        \end{flushleft}
        \centering
            \includegraphics[width=\linewidth]{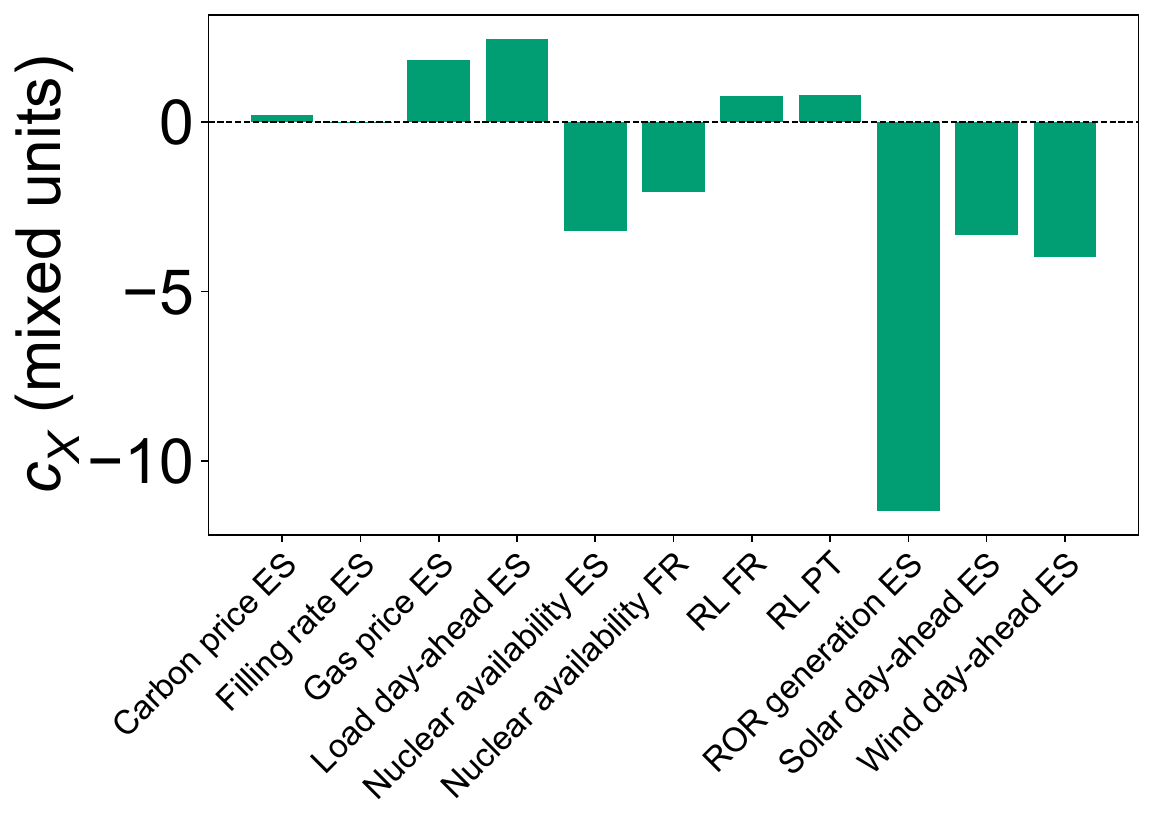}   
    \end{minipage}
    \caption{
    \textbf{Coefficients of SCM.}
    Coefficients of SCM with electricity market price of France \textbf{a}, net exports of France \textbf{b} and electricity market price of Spain \textbf{c} as target. 
    \label{fig:coefficients}}
\end{figure*}

The raw structural coefficients of the SCM models are shown in Supplementary Fig.~\ref{fig:coefficients}. It is visible that French and Spanish Nuclear availability have similar impact on the French and Spanish electricity market prices.
The units are mixed, for the electricity market prices units are given by $ MWh \cdot tCO2e $ for Carbon price, unit less for gas price, $EUR/MWh \cdot GWh^{-1}$ and in $EUR/MWh \cdot GW^{-1}$ for the rest. For the net exports, units are given by $100 \, tCO2e \cdot MW/EUR$ for the Carbon price, $100 \, MW^2 h/EUR$ for the gas price, $1/h$ for filling rate and unit less for the rest.

\refstepcounter{notecounter}
\subsection*{Supplementary Note \thenotecounter:~Share of natural gas generation in french total generation}

France relies mostly on nuclear power for its electricity generation. Compared to its neighbours, gas generation plays a lesser role in French electricity production. In Supplementary Fig.~\ref{fig:gas generation} the share of gas generation of the total french generation is shown. During the energy crisis in the year 2022, an increase of share of gas generation can be observed.

\begin{figure}[tb]
    \includegraphics[width=\linewidth]{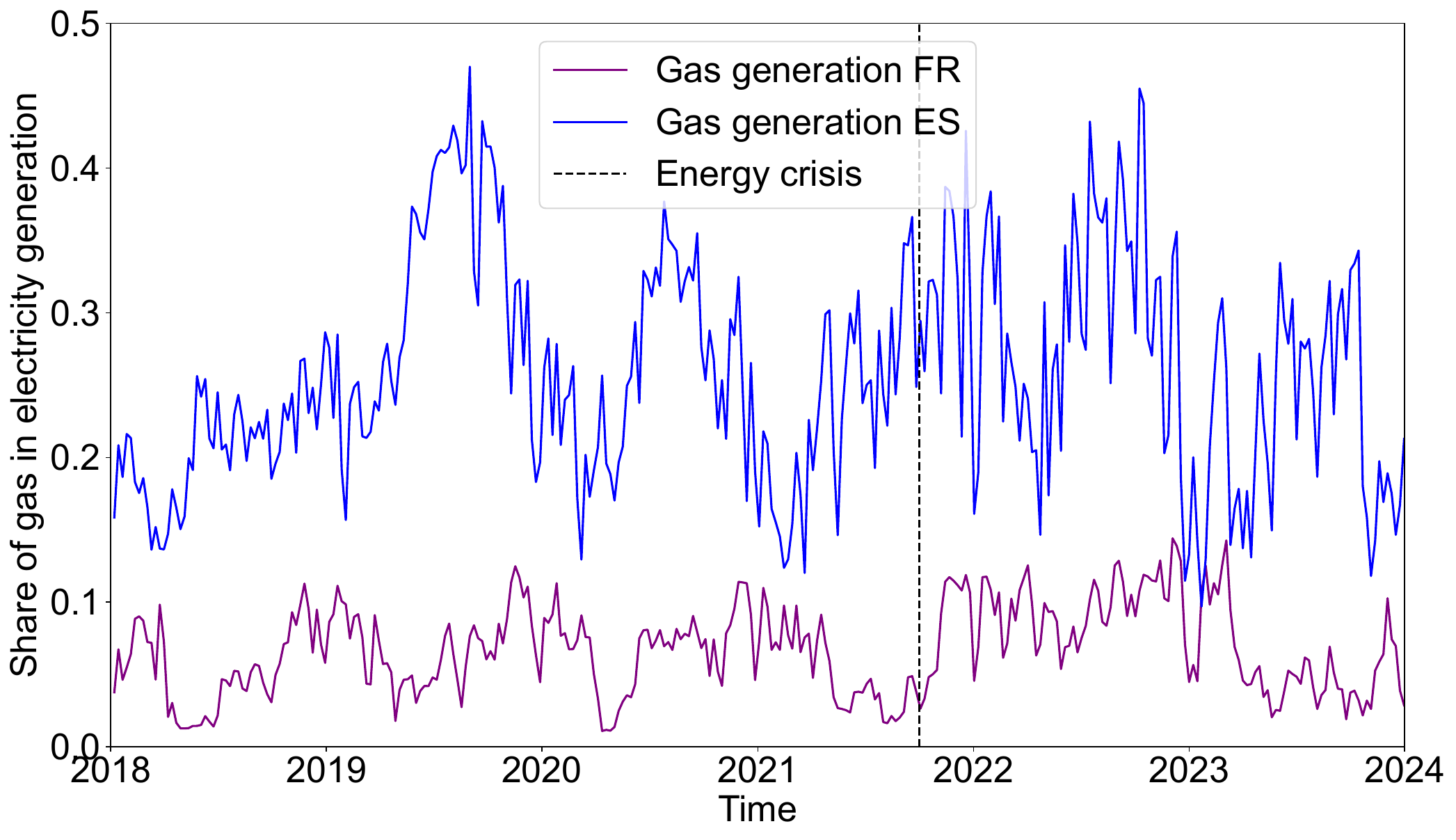}          
    \caption{\textbf{Share of natural gas in french electricity generation.} 
    We plot the weekly average share of gas generation of the total french and Spanish electricity generation. The
    beginning of the energy crisis defined as 1st October 2021 is marked with a dotted line. The share of gas generation increases after the start of the energy crisis in the year 2022.}
    \label{fig:gas generation}
\end{figure}

\refstepcounter{notecounter}
\subsection*{Supplementary Note \thenotecounter:~Price differences between France and its neighbours}

We use SCMs shown to analyse why French electricity market prices strongly increased during the energy crisis even though France relies less on gas than its neighbours. We conclude that two factors caused the sharp rise in French electricity market prices: The unavailability of French nuclear power plants leads to France being more dependent on its neighbours. This leads to higher French electricity market prices during the energy crisis when compared with its neighbours~\cite{banque2023energy}.

\begin{figure}[tb]
    \includegraphics[width=\linewidth]{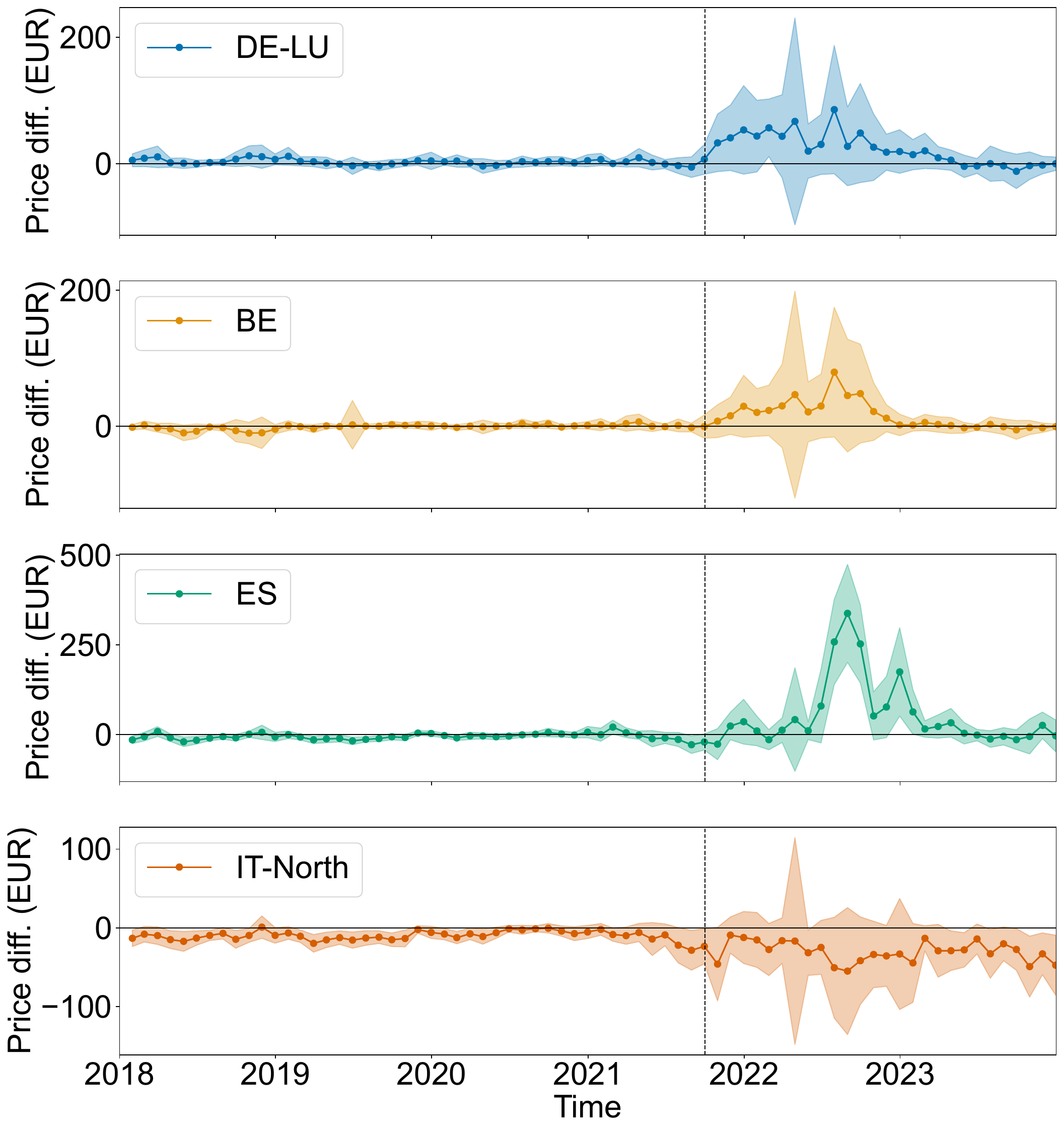}          
    \caption{\textbf{Differences of French electricity market prices with respect to the neighbouring bidding zones.} 
    We provide a monthly aggregated statistics, where the circles give the mean and the shaded area in
    indicates the standard deviation. The solid line is drawn to guide the eye. The beginning of the energy crisis defined as 1st October 2021 is marked with a dotted line.}
    \label{fig:pricesync}
\end{figure}

We analyse the difference of electricity market prices in France with respect to the neighbouring bidding zones in Supplementary Fig.~\ref{fig:pricesync}. Before the energy crisis, price differences with respect to Spain (ES), Belgium (BE) and Germany-Luxembourg (DE-LU) are generally small. They are mostly negative for ES and vary between positive and negative values for BE and DE-LU. 
During the crisis, price differences with respect to these neighbours increase sharply. Positive values are much more frequent than negative values, indicating imports from the respective bidding zones to France. Hence, the French electricity market could not isolate itself from its neighbours, which have a substantial share of natural gas in the electricity mix. 

\begin{figure*}[tb]
    \includegraphics[width = .8\linewidth]{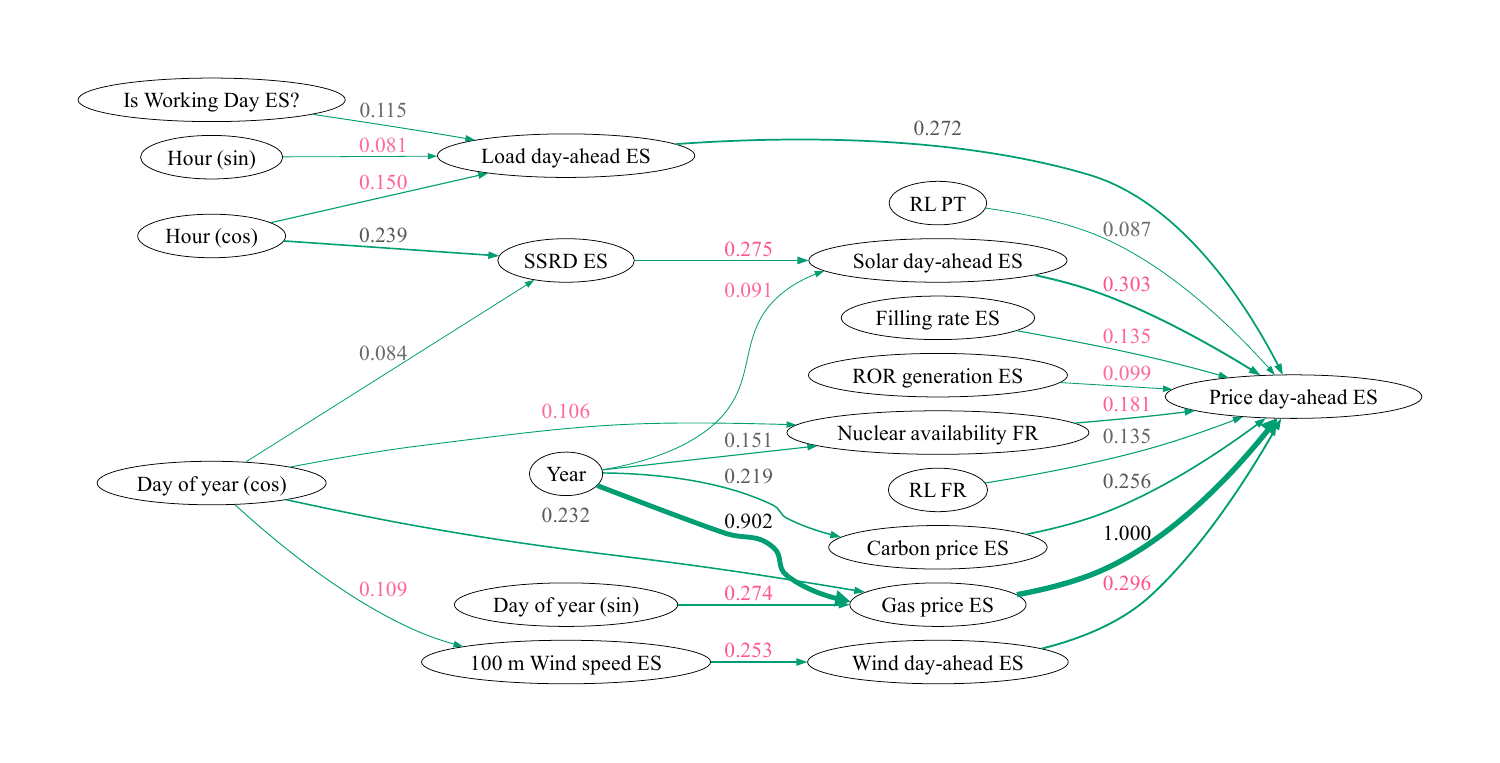}
    \caption{
    \textbf{Mean absolute Shapley Flow values explain the non-linear GBT model for the electricity market prices in Spain.} Only the 25 most important edges are shown. The edge attributions quantify how a given feature directly or indirectly influences the target prediction and is given in  \si{EUR/MWh}. The color indicates a positive (grey) or negative (red) correlation coefficient between Shapley Flow values on the respective edge and the feature value of the starting node.
    }
    \label{fig:causal_graph_flow_spain}
\end{figure*}

The situation is different for the bidding zone IT-North. Italy frequently imports electric power from its neighbours~\cite{trebbien2024patterns} and often sees comparably high electricity market prices. 
Accordingly, price differences between France and IT-North are often negative. During the energy crisis, the monthly average of the price difference remained negative while the standard deviation increased strongly. 

Overall, the correlations of electricity market prices with Italy increased during the energy crisis~\cite{trebbien2024patterns} contrary to the other neighbours. That is, electricity market prices in France showed similar patterns it Italy, where natural gas provides contributes substantially to electricity generation (see main text).

\refstepcounter{notecounter}
\subsection*{Supplementary Note \thenotecounter:~Performance of GBT models}
\label{sec:GBT-performance}

We complement the usage of SCMs with GBTs, as these can model non-linear patterns in the data. 
Hence, we also observe a better performance of the GBT models compared to the SCM models. We report $R^2$ scores and other metrics in Supplementary Table~\ref{tab:model_performance} and Supplementary Fig.~\ref{fig:R2 scores} for the SCM and in Supplementary Table~\ref{tab:model_performance_gbt} and Supplementary Fig.~\ref{fig:r2-scores_gbt} for the GBT model.  
Due to the non-linearity of GBT models, it is possible to substantially increase the $R^2$ score and thus the explained variance, particularly for features that cannot be adequately explained by linear SCMs. 
While all targets achieve an $R^2$ score of at least 92 percent, all non-root nodes in the causal graph have an $R^2$ score of at least 50 percent, and the majority have a score of over 70 percent. This confirms that GBT models are useful for analysing indirect effects in causal graphs and complement the results of SCMs in a meaningful way.

\begin{figure*}[tb]
    \centering  
    \includegraphics[width = .8\linewidth]{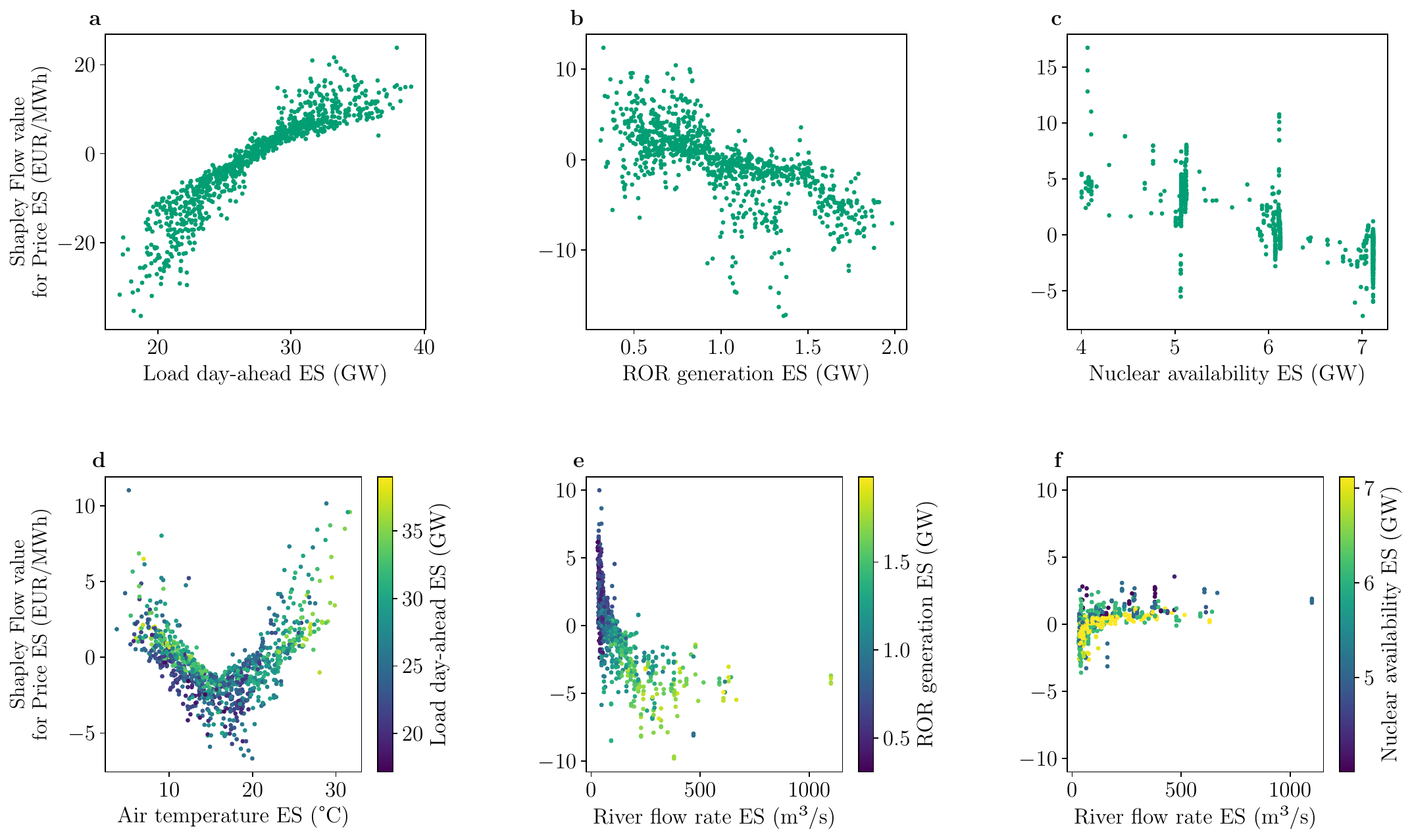}
    \caption{
    \textbf{Shapley Flow analysis reveals direct linear and indirect non-linear dependencies.}  
    \textbf{a-c,} Shapley Flow dependencies plots for direct effects in the model for the Spanish electricity market price. 
    \textbf{d-f,} Shapley Flow dependencies plots for indirect effects on the electricity market price. 
    Air temperature via Load, River flow rate via ROR generation and River flow rate via Nuclear availability. The value of the intermediate variable is colour coded. 
    }
    \label{fig:dependence_plot_prices_spain}
\end{figure*}

\refstepcounter{notecounter}
\subsection*{Supplementary Note \thenotecounter:~Further Shapley Flow results}

Here we provide additional results for Shapley Flow in the GBT model. This analysis complements the results of the models of the electricity market prices and net exports of France in the main text.

An overview of the Shapley Flow values in the model of the electricity market price in Spain is provided in Supplementary Fig.~\ref{fig:causal_graph_flow_spain}. As in the linear SCM, the gas price is the most important feature, complemented by the day-ahead wind and solar generation. 
These two features have larger impact for the Spanish electricity market prices than in the model for the electricity market prices in France. 
However, the type of correlation between the feature values pointing to the target and the Shapley flow values of the corresponding edges mostly resembles that of the French model. 

We continue with a analysis of linear and non-linear effects of meteorological and economical features for the electricity market price in Spain. When analysing the direct effects of day-ahead load, ROR generation, and nuclear availability, we see similar effects on electricity market price in Spain as in France, see Supplementary Fig.~\ref{fig:dependence_plot_prices_spain}a-c. 
With regard to the indirect effects of air temperature, we see similar effects on the price via the day-ahead load as in France, up to a temperature of \qty{20}{\degreeCelsius} in panel d. Above \qty{20}{\degreeCelsius}, we see a more pronounced increase, which is probably due to increased use of air conditioning in Spain, compared to France.

In addition to the most relevant graph edges with respect to the Shapley Flow values, as discussed in the main text and in Supplementary Fig.~\ref{fig:causal_graph_flow_spain}, we show for each of the three models a sub-graph with all nodes and edges (and thereby all flows). Here we focus on the temperature impact, specifically the indirect effects of air temperature, river temperature, and river flow rate on the model target, see Supplementary Fig. \ref{fig:zoom_in_causal_graphs}. 
We note that each of these three features has a substantial impact on prices and net exports via both nuclear availability and run-of-river generation.

While the indirect effects of river flow rate on ROR generation are similar for all three models, the impact via nuclear availability in Spain is low in comparison.

To model the impact of hypothetical gas prices in Spain and nuclear availability in France during the energy crisis, we use a what-if scenario in which we apply the trained GBT models for predictions with hypothetical input values.
In Supplementary Fig.~\ref{fig:what-if_shapley_flow}, we see that the Shapley flow values for both hypothetical gas prices as well as nuclear availability are in a similar range to the values for empirical values from the model's test set. Hence, we are not prone to extrapolate to completely unknown feature values. We conclude that the predictive power of the models for hypothetical gas prices and nuclear availability is reliable.

\refstepcounter{methodcounter}
\subsection*{Supplementary Methods:~Data sources and pre-processing}

Data is obtained and processed as explained in the main text. Here we provide additional information.

Multiple bidding zone changes occurred during the period of investigation. Austria left a shared bidding zone with Germany and Luxembourg in 2018. The time series data of the bidding zone DE-AT-LU (until 2018-09-30) are therefore combined with the one of DE-LU (from 2018-10-01 onwards).
The Italian bidding zones were modified on 2019-01-01 and on 2021-01-01~\cite{ternaNuoveZone}. 
The production hubs IT-Brindisi, IT-Foggia and IT-Priolo were removed on 2019-01-01. On 2021-01-01, the production hub IT-Rossano was removed and the bidding zone IT-Calabria added. Meanwhile the region Umbria moved from IT-Centre-North to IT-Centre-South. The data for the production hubs before removal and the data of IT-Calabria after its addition is aggregated to the bidding zone of IT-South.
The bidding zone IT-North, that is considered in the causal model, was not affected.

The aggregated net exports were calculated using the total scheduled exchange considering all neighbours. That is, we add the exports to all neighbouring bidding zones and subtract the imports to all neighbouring bidding zones. We note that we use the total scheduled exchanges instead of the day-ahead scheduled commercial exchange for data quality reasons. In particular, the share of missing data is higher for day-ahead values.
In addition Spain has the non-SDAC neighbours Morocco and Andorra. Data of these exchanges can be found using the API of Red Électrica \cite{REE_API}.

The residual load is calculated by subtracting day-ahead solar and wind generation from the day-ahead load. ROR generation is hereby neglected to avoid mixing day-ahead and actual generation data.

Since there have been some inconsistencies in ENTSO-E data of unavailable production and generation since the update of the website in the end of 2025, we use alternative sources. In case of France, data on unavailability of nuclear power plants was taken from RTE \cite{RTE_IIP_production_unavailability}. Here,
we only consider unavailability due to planned maintenance and
outages longer than 24 hours, since shorter outages are assumed to
not affect the day-ahead price. The nuclear availability is obtained by subtracting the unavailable
nuclear capacity from the total installed nuclear capacity. Data on installed nuclear capacity of France is retrieved from ENTSO-E manually with permanent links for reproducability \cite{ENTSOE_nuclear_cap_2018_2023}.
In case of Spain ESIOS of Red Électrica \cite{REE_ESIOS_analysis} provides data on accumulated nuclear availability directly, thus not allowing for differentiation between planned and unplanned maintenance.
Data of filling rate of Hydro Storage plants and reservoirs is downloaded manually from ENTSO-E manually with permanent links for reproducability \cite{ENTSOE_filling_rate_FR_2018_2023, ENTSOE_filling_rate_ES_2018_2023}.The data is available in weekly resolution and is padded to hourly resolution.

Weather data including surface solar radiation downwards, wind speed at \qty{100}{\meter} and air temperature at \qty{2}{\meter} is provided by ERA5 \cite{C3S_ERA5_2023, Hersbach_ERA5_2023}, mean values are calculated by averaging over larger areas including France and Spain respectively as shown in Supplementary Fig. \ref{fig:french grid}. The considered area for France is bounded by latitudes \qty{41}{\degree}-\qty{52}{\degree} $N$ and longitudes \qty{6}{\degree} $W$-\qty{10}{\degree} $E$ , and latitudes \qty{36}{\degree}-\qty{44}{\degree} $N$ and longitudes \qty{10}{\degree} $W$-\qty{5}{\degree} $E$ in case of Spain.

The API Hub'Eau~\cite{eaufranceAccueilHubeau} provides hourly river temperature data and daily river flow rate data of France. We consider rivers with the highest mean flow rates and those near nuclear power plants: Rhine, Meuse, Rh\^{o}ne, Loire, Garonne, Seine, Dordogne, Vienne and Moselle. For each river, we averaged the data from different measuring stations. Then, we averaged both variables over all rivers.
In case of Spain, we used daily river temperature and flow data of the river Ebro as a proxy. The data is provided by SAIH Ebro \cite{CHE_SAIH_Ebro_historicos}.

All time series used in the analysis are denoted in coordinated universal time (UTC) and have an hourly resolution. Some time series, as for example the gas price, are available only with daily resolution. In this case we padded the daily value for all hours of the day.

\begin{figure*}[tb]
    \begin{flushleft} 
        \hspace{3cm}\textbf{a}
    \end{flushleft}
    \vspace{-1cm}
    \includegraphics[width = \linewidth]{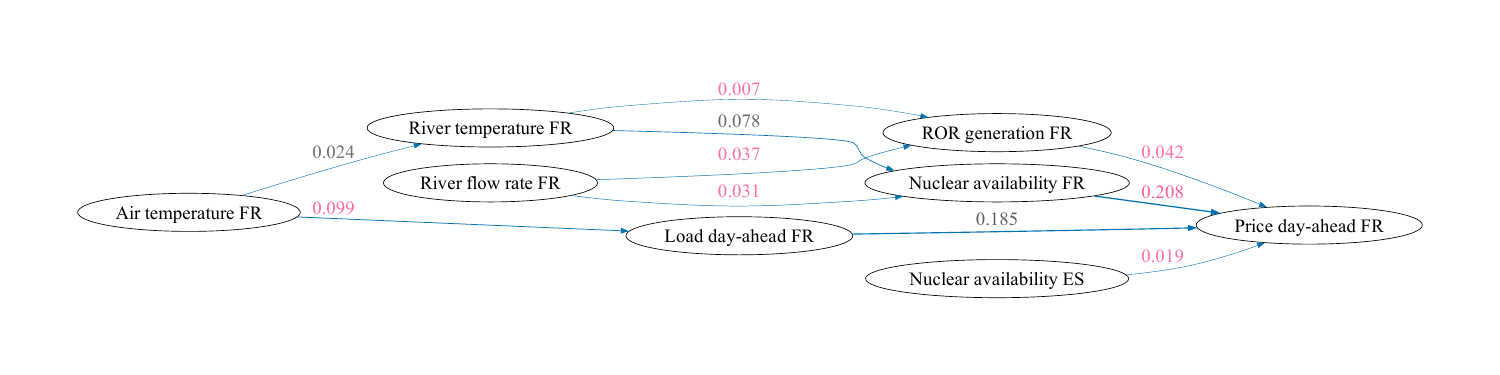}
    \begin{flushleft} 
        \hspace{3cm}\textbf{b}
    \end{flushleft}
    \vspace{-1cm}
    \includegraphics[width = \linewidth]{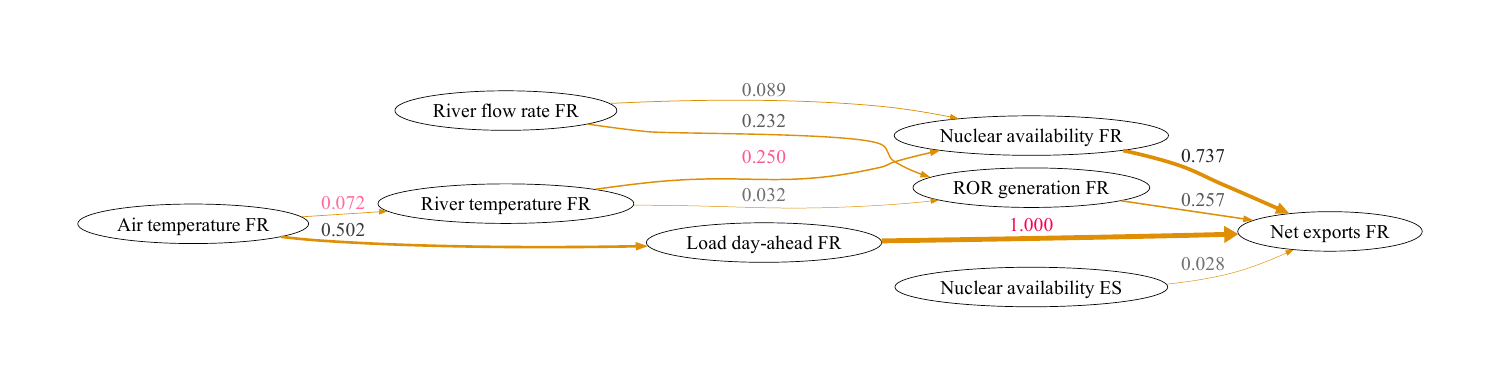}
    \begin{flushleft} 
        \hspace{3cm}\textbf{c}
    \end{flushleft}
    \vspace{-1cm}
    \includegraphics[width = \linewidth]{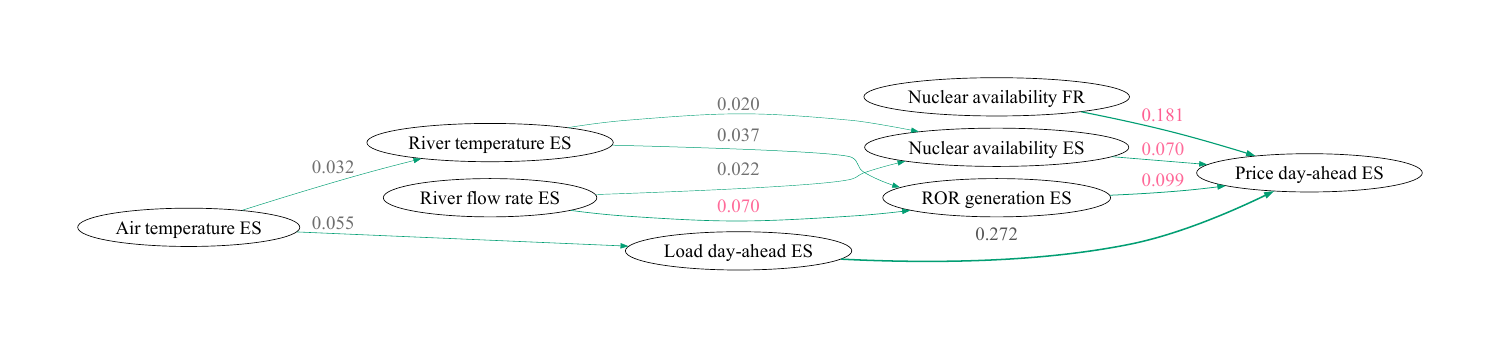}
    \caption{
    \textbf{Mean absolute Shapley Flow values for a sub-graph of the causal graph.}
    \textbf{a,} Results for the electricity market price in France.
    \textbf{b,} Results for the net exports of France.
    \textbf{c,} Results for the electricity market price in Spain.
    We focus on the indirect effects of air temperature, river temperature and river flow rate on the respective target. 
    }
    \label{fig:zoom_in_causal_graphs}
\end{figure*}

\begin{figure*}[tb]
    \centering
    \includegraphics[width = \linewidth]{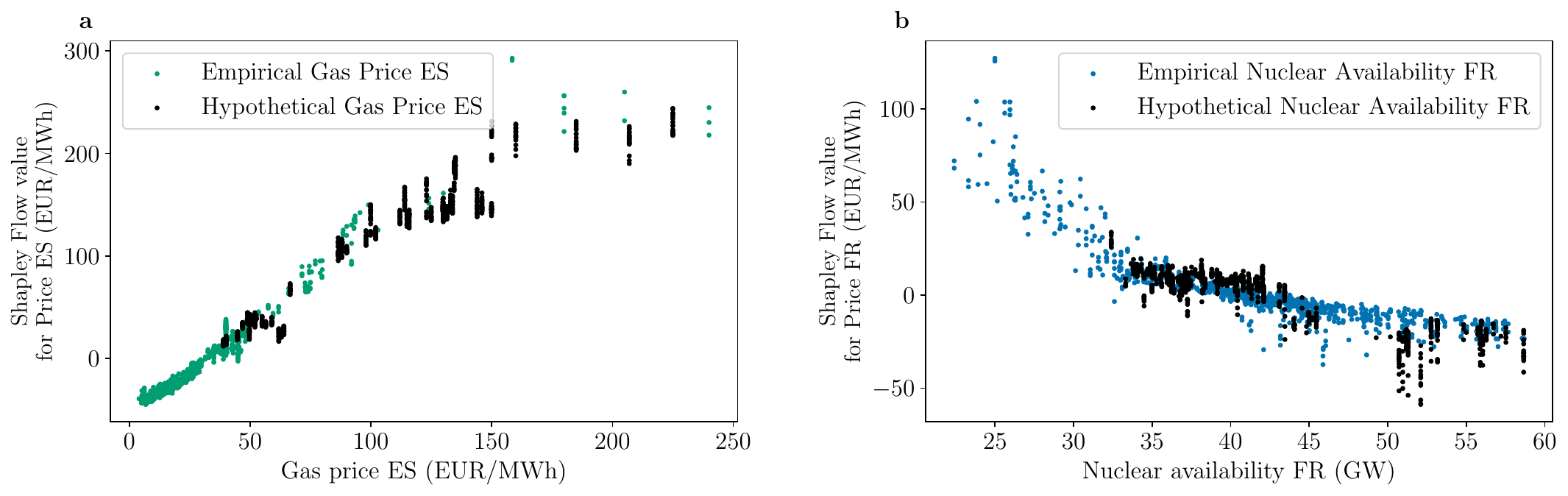}
    \caption{
    \textbf{Shapley Flow values for the What-if scenarios.} 
    \textbf{a,} Spanish electricity market price with the hypothetical scenario without a gas price cap.
    \textbf{b,} French electricity market price with the hypothetical scenario of \SI{10}{GW} additional nuclear availability.
    We compare Shapley Flow values for the model with original empirical input data (green/blue dots) and hypothetical input data (black). 
    We note that in both what-if scenarios, the value ranges of the Shapley flow values of the hypothetical samples correspond to the ranges of the empirical samples.}
    \label{fig:what-if_shapley_flow}
\end{figure*}

\begin{figure*}[tb]
    \centering
    \begin{minipage}{0.45\textwidth} 
        \centering
            \includegraphics[width=\linewidth]{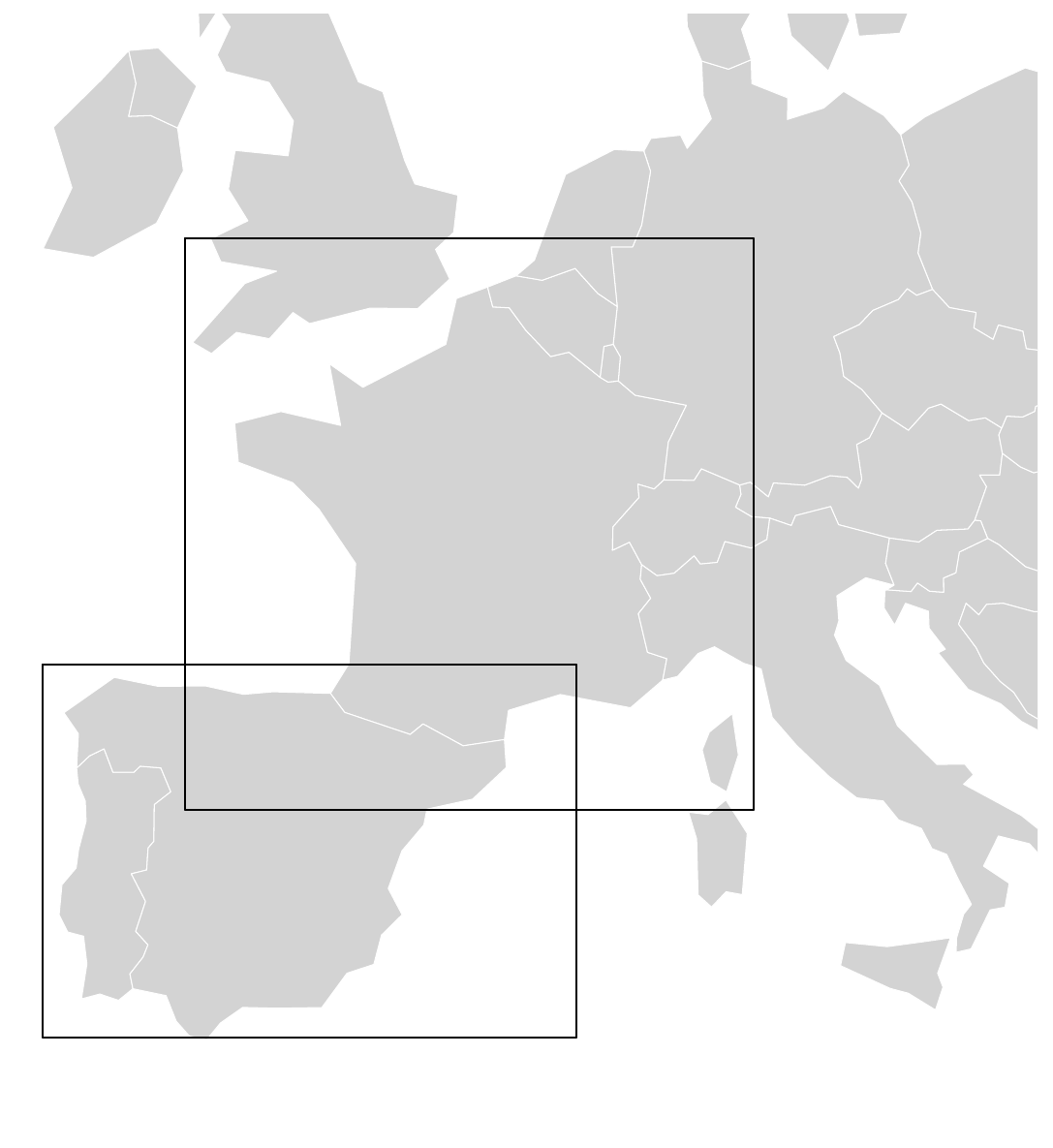}       
    \end{minipage}
    \caption{\textbf{Definition of the weather data.} Weather data is gathered from ERA 5. The areas of France and Spain are hereby approximated with the shown rectangles. Country borders are based on Natural Earth shapefiles \cite{naturalearth}.}
    \label{fig:french grid}
\end{figure*}

\clearpage


\end{document}